\documentclass[utf8]{frontiersinFPHY_FAMS} 
\usepackage{subfig}
\usepackage{url,hyperref,lineno,microtype}
\usepackage[onehalfspacing]{setspace}

\def\keyFont{\fontsize{8}{11}\helveticabold }
\def\firstAuthorLast{Dijcks et al} 
\def\Authors{Siebe Dijcks$^1$†, Lukáš Kusýn$^{2, 1}$†, Jesper Janssen$^{3,1}$, Petr Bílek$^4$, Sander Nijdam$^1$‡, Tomáš Hoder$^2$‡} 
\def\Address{$^1$Eindhoven University of Technology, Eindhoven, The Netherlands\\
$^2$Department of Physical Electronics, Faculty of Science, Masaryk University, Brno, Czech Republic\\
$^3$Plasma Matters BV, Eindhoven, The Netherlands\\
$^4$Institute of Plasma Physics of the Czech Academy of Sciences, Prague, Czech Republic}

\def\firstA{†These authors contributed equally to this work and share first authorship\\
‡These authors contributed equally to this work and share last authorship}

\def\corrAuthor{Sander Nijdam}

\def\corrEmail{s.nijdam@tue.nl}

\begin{document}
\onecolumn
\firstpage{1}

\title[E-field and temperature distributions in positive streamers]{High-resolution electric field and temperature distributions in positive streamers}

\author[\firstAuthorLast ]{\Authors} 
\address{} 
\FirstA{} 
\correspondance{} 

\extraAuth{}

\maketitle

\begin{abstract}

\section{}
In this work, we aim to take a detailed experimental picture of the positive streamer.
We apply optical emission spectroscopy to the first negative system (FNS, ${\rm B}^2\Sigma_{\rm u}^+\rightarrow {\rm X}^2\Sigma_{\rm g}^+$) of N$_2^+$ and the second positive system (SPS, ${\rm C}^3\Pi_{\rm u}\rightarrow {\rm B}^3\Pi_{\rm g}$) of N$_2$.
Large, centimeter wide, and highly reproducible streamers are created in pure nitrogen and synthetic air, at pressures ranging from 33 to 266\,mbar.
Direct time resolved spectral imaging of the space charge layer resulted in spatiotemporal maps of the calculated reduced electric field strength ($E/N$) and rovibrational temperature in sub-nanosecond and sub-millimetre resolution.
The $E/N$ peaks at approximately 540 and 480\,Td, directly in front of the space charge layer, for synthetic air and pure nitrogen respectively, as determined by using the intensity ratio method of FNS and SPS.
A global model for pure nitrogen in PLASIMO uses the experimentally determined $E/N$ distribution to draw a picture of the gas kinetics around the space charge layer passage.
In addition, the results of the global model serve as a reference to interpret the rotational and vibrational temperatures obtained from experimental FNS and SPS emissions, which emphasize the streamers' non-equilibrium gas heating.

\tiny
 \keyFont{ \section{Keywords:} plasma, streamer discharge, reduced electric field, optical emission spectroscopy, OES, nanosecond pulsed discharge, non-equilibrium} 
\end{abstract}

\section{Introduction}
Streamers are a discharge type that can develop when strong local electric fields are applied to a gaseous medium. 
The paths of these growing and ionizing ‘fingers’ are paved by a space charge layer, locally enhancing the electric field strength, ionizing the species ahead of this layer and thereby propagating the streamer~\cite{Nijdam2020}. 
Generally, these streamer discharges develop into complex morphological structures due to the stochastic branching they experience while propagating.

Since streamers can be a precursor to short-circuits, a streamer-initiated breakdown~\cite{cernak2020}, they are undesirable in high voltage engineering and therefore something to actively avoid~\cite{Seeger2018}.
However, the non-equilibrium nature of streamers and their strong local electric fields make them of interest in various gas treatment applications~\cite{Adamovich2022}.
Example applications range from creating high energy species like ozone in air purification plants to nitrogen fixation in plasma activated water where it acts as a fertilizer.
Basic experiments like in this work where streamer morphologies and energetics are studied in tandem are valuable for validating modelling efforts. 

Two of the main parameters of interest when characterizing streamer discharges are the reduced local electric field strength ($E/N$) and the electron energy distribution function (EEDF) which describe the electrons accelerated by this field.
There are methods of measuring these properties directly and indirectly for plasmas and streamer like discharges~\cite{Goldberg2022,Laux2003,Simek2014,Engeln2020}.
However, it has been proven very difficult to apply these measurement techniques to transient and stochastic phenomena like a streamer discharge, which is exacerbated by the low active species densities found in these types of plasmas.
A method to measure the electric field which has become popular in recent years is Electric Field Induced Second Harmonic generation (E-FISH). However, this method suffers from complications which make it very hard to use for the quantitative determination of electric fields with unknown spatial distributions~\cite{Goldberg2022,Chng2022}.

In this work, the method of the intensity ratio of the FNS and SPS optical emissions is applied to quantify the electric field in the streamer head. 
The method was typically used in millimetre sized barrier discharges and short gaps at atmospheric pressure, see e.g. the pioneering works of Kozlov, Shcherbakov and their coworkers~\cite{kozlov2001spatio,shcherbakov2007_2}.
Recently Lepikhin et al~\cite{Lepikhin_2022} have applied the same method for the determination of reduced electric field with a comparison of the field from a capacitive probe, demonstrating a significant discrepancy of the reduced electric field in the first few nanoseconds of the discharge. 
The work is focused on nanosecond capillary discharge at 27\,mbar and discusses the importance of detailed investigation of the temporal development of used FNS and SPS emission bands and population/depopulation channels of their upper states. 
A similar analysis was done using the uncertainty quantification and sensitivity analysis methods by Obrusník, Bílek and coworkers~\cite{obrusnik2018electric,bilek2018electric}.

In this contribution, we enhance the experimental analysis further and achieve a unique spatio-temporal high-resolution insight into the electric field distribution in the streamer head by using optical emission spectroscopy (OES).
This is achieved, firstly, by designing the electrostatic electric field distribution in the discharge gap, such that streamers cross the electrode gap on cue without branching in a very repetitive manner.
And secondly, by performing the streamer discharge experiments at reduced pressures, which gives relatively large, centimeter sized, cylindrically symmetric single streamer filaments.
This allows for spatially and temporally spectral imaging of the streamers' optical emission fingerprint, from which the reduced electric field is calculated.
The added advantage of experimenting at reduced pressure is that the observer is essentially zooming in, both spatially and temporally, to the streamer discharge, following the scaling laws described by Ebert et al~\cite{Ebert2010}.
Our reduced electric field profiles determined for synthetic air are compared with numerical results obtained by Babaeva and Naidis~\cite{Babaeva2021} and Li et al~\cite{Li2021}.
Moreover, we apply a spectral fitting procedure to obtain the distributions of rotational and vibrational temperatures in the streamer head and tail with similar resolution, which is directly compared to the numerical simulation presented in this work.
Comparison to the global model enables a better insight into the processes taking place around the streamer head passage. 

In the next sections, the experimental setup and methods are described in detail as used for the streamer analysis. 
In the fourth section, the results are presented and discussed. 
The conclusions are presented in the final fifth section.

\section{Experimental setup} 
\begin{figure}
\begin{center}
 \includegraphics[width=0.8\textwidth]{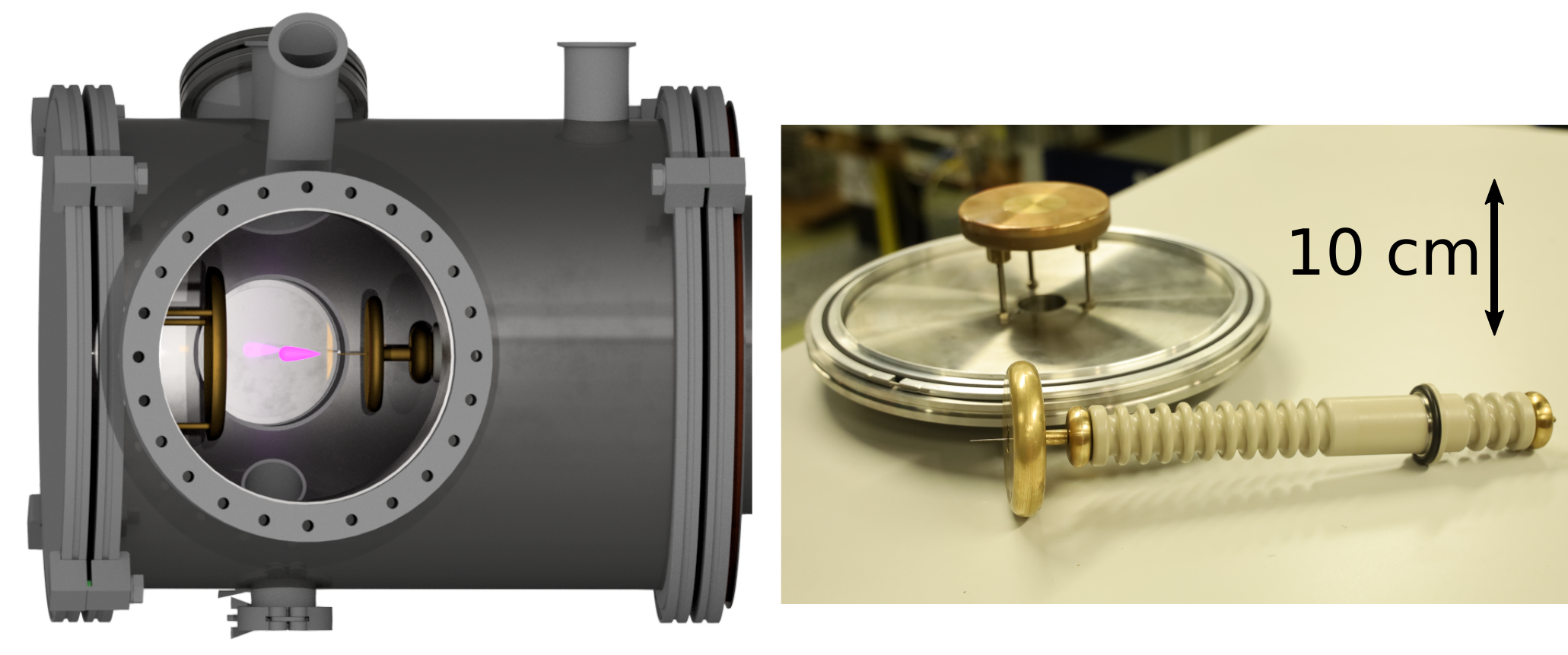}
\end{center}
\caption{On the left a 3D render of the discharge vessel, including a representation of a streamer propagating through the discharge gap, shown propagating from right to left. The image is rendered from the viewpoint of the monochromator entrance slit and ICCD camera, viewing the discharge and electrodes through a large quartz window. On the right the unmounted electrodes, with the grounded electrode screwed to the vessel flange and the powered electrode with rod and PEEK KF\,40 feed though.}
\label{3D_render}
\end{figure}

The setup revolves around the discharge volume where the streamers propagate in a repetitive manner.
The identical repetitive streamers make the phase resolved image sequence, synchronised to the voltage waveform, into a time resolved sequence. 
Figure~\ref{3D_render} depicts the cylindrical vacuum vessel, with inside the electrodes between which the streamers propagate.
A large electromagnetic compatibility (EMC) box is attached to the vessel flange on the right, which contains the electronic circuit producing the high voltage pulses.
The powered electrode is directly connected to the pulse source via the feedthrough pictured on the right side of figure~\ref{3D_render}.

\begin{figure}[h!]%
    \centering
    \subfloat[ ]{{
    \label{phase_resolved}
    \includegraphics[width=0.51\textwidth]{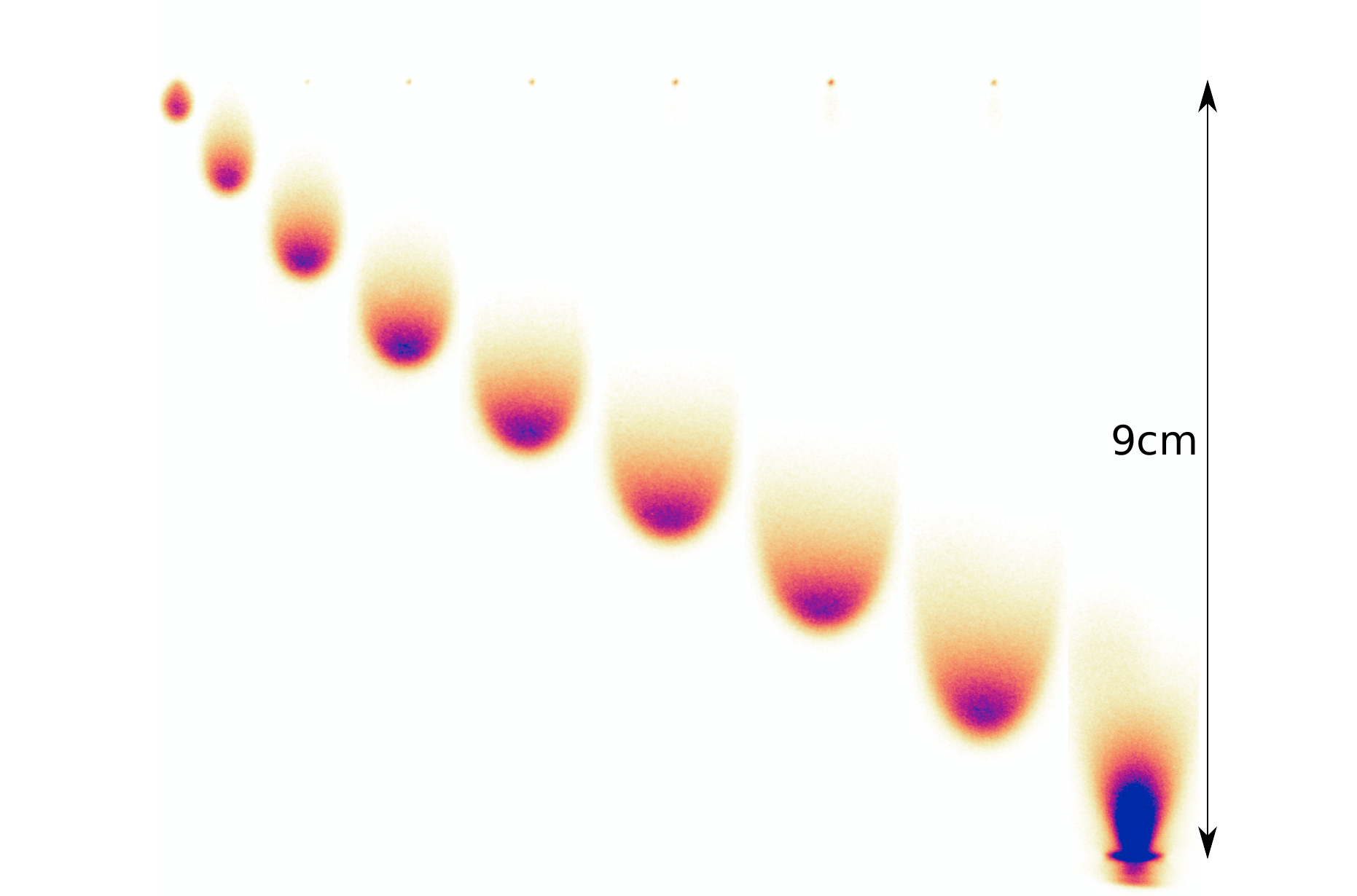}}}%
    \qquad
    \subfloat[ ]{
    \label{ratios_streamers}
    {\includegraphics[width=0.41\textwidth]{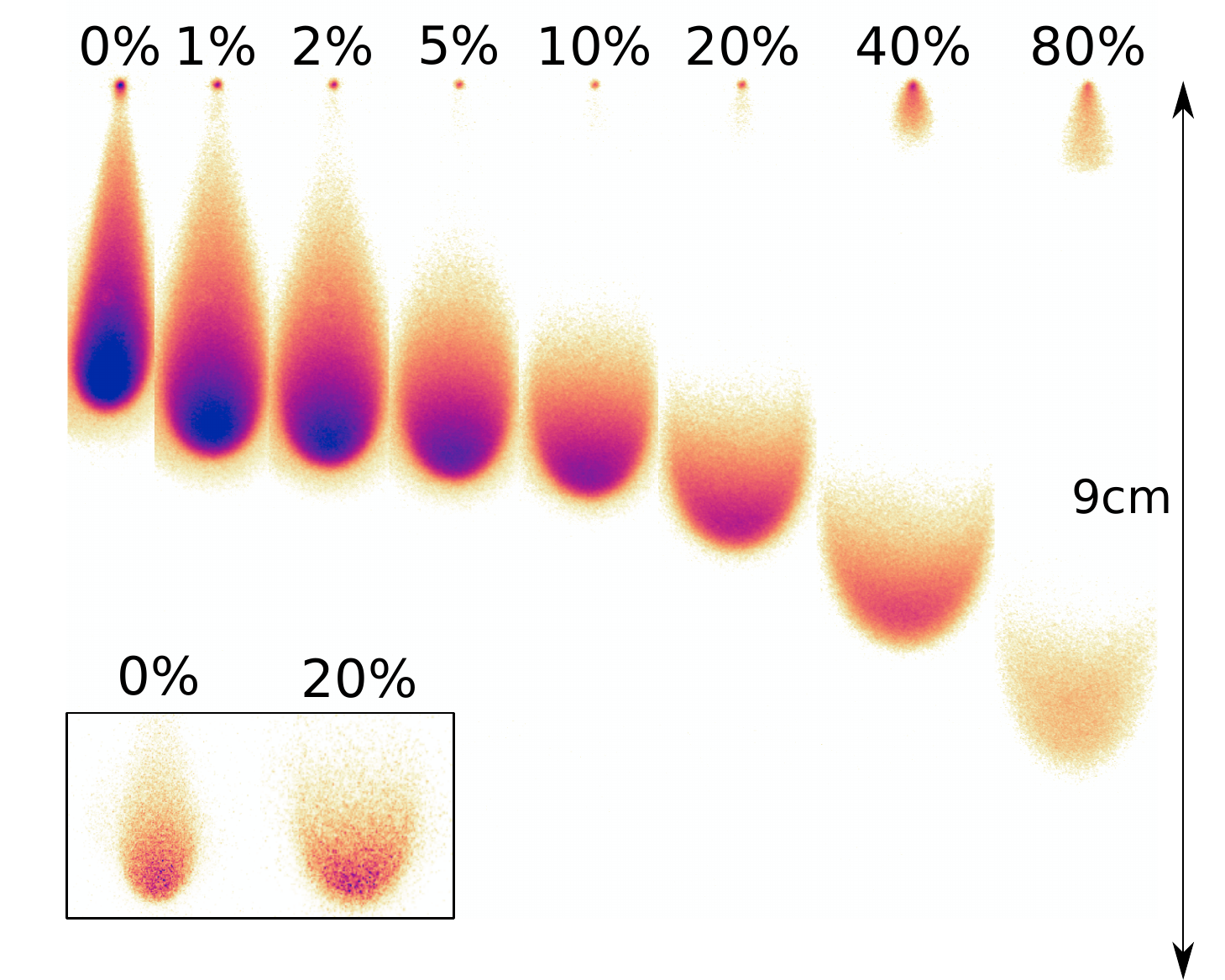} }}%
    \centering
    \caption{a) False colour images with linear intensity scaling of the phase resolved downward propagation of single streamers in synthetic air at 66\,mbar and 8\,kV. The side by side exposures are cut from the original image to only show the streamer area. The time between images is 20\,ns and each image has a 2\,ns gate. b) False colour images with logarithmic intensity scaling of streamers crossing the gap at various N$_2$/O$_2$ ratios, where the number at the top describes the percentage of oxygen admixed to pure nitrogen. All images are taken in 66\,mbar and 8\,kV at identical delays related to the voltage pulse rise and all images consist out of 50 accumulations, except for the images in the inset, which depict single shots for the 0 and 20\,\% oxygen.}%
    \label{streamer_prop}%
\end{figure}

The experimental setup is designed to create repetitive single streamer discharges at reduced pressures, in this work from 33 to 266\,mbar, for synthetic air and pure nitrogen.
At reduced pressure, the spatial length scales are blown up and temporal scales are stretched due to the increased mean free path and reduced collision frequencies at reduced pressures.
This effectively allows for zooming in to the streamer discharge~\cite{Ebert2010}.
Since spatial and temporal resolutions are experimentally fixed by the used equipment (Imaging lenses and intensifier gating), the increased size results in a relatively higher resolution when studying the transient and filamentary streamers.
The filaments of the streamer created with this setup at 8\,kV and 33\,mbar are optically around 1\,cm thick, see figure~\ref{phase_resolved}.
The space charge layer is imaged spectrally resolved with a time resolution of 500\,ps while it propagates the gap with a velocity of around $5\cdot10^5$\,m/s.

\begin{figure}
    \centering
 \includegraphics[width=1\textwidth]{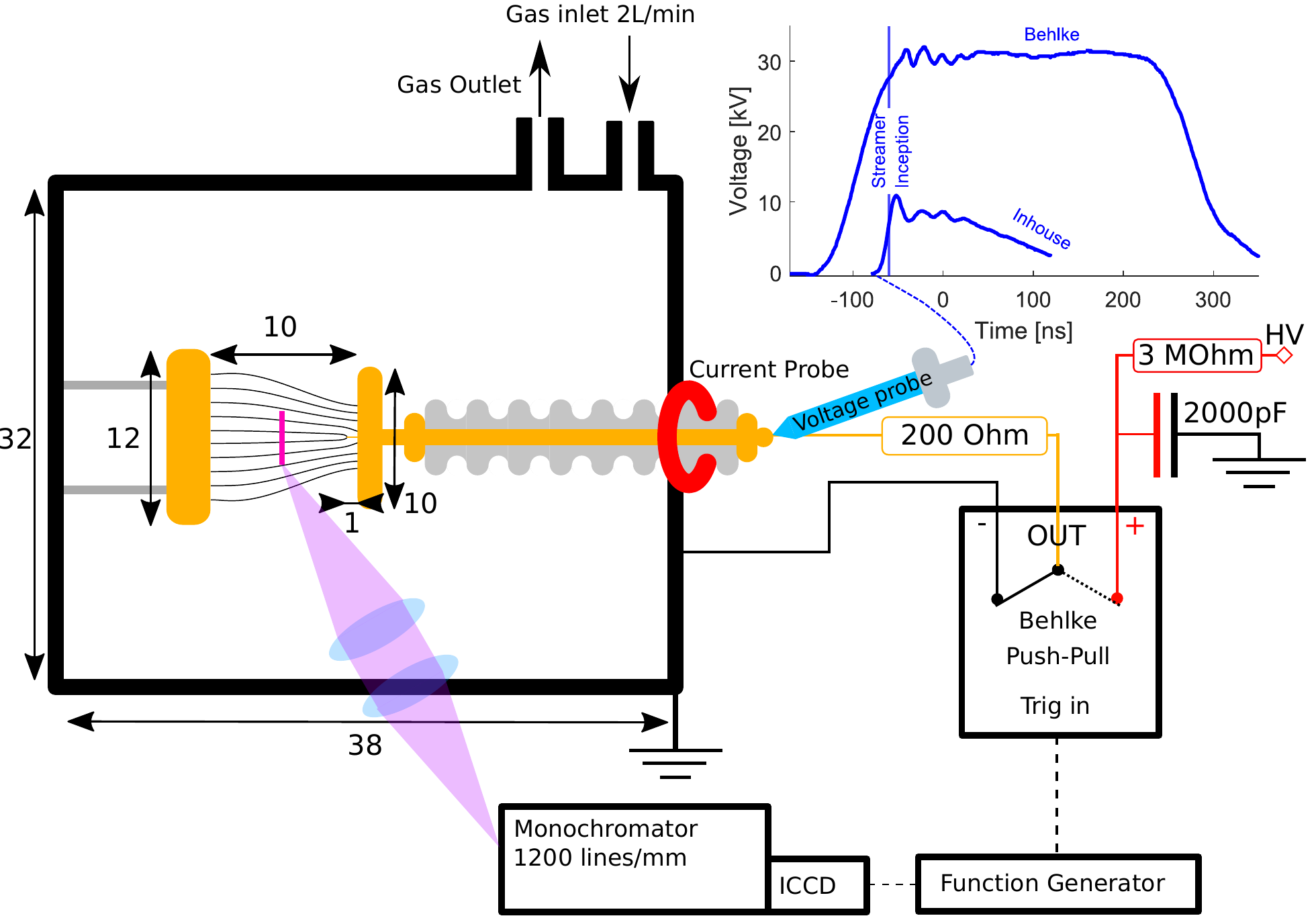}
\caption{Schematic overview of setup, with dimensions in centimeters. Between the opposing disk electrodes calculated electrostatic field lines of the field before streamer inception are drawn. The magenta bar shows the region which is imaged on the monochromator entrance slit. The voltage waveforms for the two pulse circuits are shown, and the time axis is synchronized with figure~\ref{33mbarAirAxial}; the actual amplitude and width are adjusted between experiments. Only one of the two circuits (the Behlke push-pull circuit) is shown.}
\label{exp_setup_schematic}
\end{figure}

The high reproducibility of the streamer discharges is achieved by placing two disk electrodes 10\,cm opposite of each other, creating a fairly homogeneous field between them, see figure~\ref{exp_setup_schematic} where the electrostatic field lines are drawn between the electrodes.
The homogeneousness of the electric field is disturbed by a short 1\,cm needle (1\,mm diameter and 60\textdegree\,tip angle) that protrudes from the powered disk electrode.
The needle serves as the origin of the spatially well-defined streamer. 
The vessel itself, with a diameter of 32\,cm is relatively far away and has no noticeable effect on the discharge.
When the diameter of the streamer is large compared to the gap distance (in this work $L_{gap}/D_{streamer}\approx~10$) the branching is effectively suppressed~\cite{Raizer2011}.
The strong field enhancement at the needle also helps to reduce the inception jitter of the streamer discharge. By using an overvolted gap, the streamer inception always occurs during the rising flank, which, when steep enough, results in low streamer inception jitter. 
The inception jitter directly affects the temporal resolution of the accumulated phase-resolved OES measurements, therefore it is essential to keep the inception and voltage waveform jitters to a minimum.
Solid state pulsers, both the Behlke push pull switch (Behlke HTS 651-15-SiC-GSM) and the custom in-house built pulser~\cite{Azizi2020} have excellent stability and repetitiveness in their output waveform.
The output voltage jitter, respectively $\leq $1\,ns and 175\,ps and the rise time, 10\,ns and 48\,ns, show that the in-house pulser gives better streamer stability.
However, the in-house pulser was limited to 10\,kV, meaning that for the pressure varying experiments, where higher voltages are required, the Belhke was used.
The high voltage source, a Spellman 205B is also very stable, with a $\leq 0.01\%$/hour stability. 

The pulse width of the voltage pulse is adjusted according to the propagation time of the streamer, such that there is no time for a streamer-to-spark transition. 
This greatly reduces the total energy input into the gas and thereby the gas heating.
For a gap distance of 9\,cm and streamer velocities of approximately $5\cdot10^5$\,m/s this means that voltage pulse widths of around 200\,ns are used.
Because the streamer has a quite fixed cylindrically symmetrical shape and a roughly constant propagation velocity in the center of the gap, a time-resolved radial measurement can be used to determine the entire 3D structure.


\setcounter{subfigure}{0}

For all the experiments the vessel is flushed with 2\,L/min, with gasses from 6.0 purity cylinders. 
The synthetic air is pre-mixed and contains 80\% nitrogen and 20\% oxygen.
Based on this the purity of the nitrogen within the vessel is estimated to be between 3.0 and 5.0.
The 0 and 20\% mixtures are premixed from a 6.0 purity cylinder, while the other ratios are manually mixed.
33\,mbar was near the lowest pressure available to the setup with 10\,cm opposing disks, taking maximum benefit of the pressure scaling, while also giving the most stable streamers in pure nitrogen.
Single streamer discharges for synthetic air and pure nitrogen under identical conditions are then possible.


The imaging and spectroscopy are performed using a Lavision PicoStar HR ICCD with UV photocathode.
For regular imaging, a 105\,mm quartz lens is used and for OES the ICCD is attached to a 1\,m Jobin Yvon monochromator with a UV-blazed 1200 lines/mm grating. 
In this work, there are two settings used to capture the spectral data for OES.
Firstly, for the high time resolution (500ps gate) acquisition, from which the $E/N$ is calculated, see figures~\ref{spectral_fits_exp},~\ref{33mbarAxial} and~\ref{fields_heatmap}, the entrance slit of the monochromator was set to 50\,$\mu$m.
Secondly, for the high spectral resolution (5ns gate) acquisition, described in figures~\ref{RadialHeatmaps} and~\ref{Plasimo_b}, the slit was set to 12\,$\mu$m.
The first setting allows for a maximum signal while sacrificing spectral detail, while for the second a spectrum with a good signal to noise ratio is obtained, which is required to perform an Abel inversion.

The central trigger comes from a function generator, sending triggers to both the pulse source and camera intensifier. 
The ICCD uses a gate and step of 500\,ps to acquire the high temporal resolution spectra.
Around 10$^5$ discharges are accumulated per time step to ensure that the signal to noise ratio is good enough for our spectral fitting methods, which at a discharge repetition rate of 200\,pps takes around ten minutes.
The streamer behaviour was verified at the beginning and the end of each approximately four hour measurement campaign and no change in inception or propagation was observed.

In figure~\ref{ratios_streamers}, the streamer crossing for various N$_2$--O$_2$ mixtures is depicted.
Higher oxygen contents lead to later streamer inception, but also to a much faster streamer propagation velocity which more than compensates for this at the time of imaging, resulting in longer streamers at the time of imaging, for higher oxygen concentration streamers.
For streamer discharges where branching does occur, the morphological structure of streamers in synthetic air versus pure nitrogen differ greatly, mainly due to the large difference in photoionization efficiency~\cite{Nijdam2010}.
High photoionization efficiency suppresses branching by supplying free electrons ahead of the space charge layer, causing many overlapping avalanches, which for positive streamers grow towards the space charge layer.
This results in smooth hemispherical streamer fronts for effectively photo-ionizing gas (mixtures), see also the single shot streamers depicted in the inset of figure~\ref{ratios_streamers}, and non-smooth `feathery' looking streamer fronts~\cite{Nijdam2010} in gas (mixtures) where photo-ionization is not a good source for free electrons. 
The single shot imaged streamers for both pure nitrogen and synthetic air show a relatively smooth front. 
For pure nitrogen this is probably due to the low pressure and small oxygen contamination.
A good estimate of the oxygen contamination in `pure' nitrogen gasses can be made by observing the streamer discharge morphology.
These contaminations cause significant photo-ionizing events, creating free electrons ahead of the streamer, resulting in observable changes in the discharge morphology.
Note that the pure nitrogen streamer is not propagating completely straight downward, but shows some curvature at the moment of imaging.
This will manifest itself in a small lateral offset when performing the spectral imaging, see the shifted center line (black dashed) for pure nitrogen in figure~\ref{fields_heatmap}.

\section{Methods}
The method described below enables us to measure and quantify the response of the gaseous medium to the strong locally enhanced electric field and estimate the reduced electric field. This is achieved by deploying the intensity ratio technique which uses the emission of specific vibronic transitions (SPS, FNS) obtained by OES measurements.

\subsection{Intensity ratio method}\label{IntensityRatioMethod}
\begin{figure}[h!]%
    \centering
    \subfloat[ ]{
    \label{el_diag}
    {\includegraphics[width=0.5\textwidth]{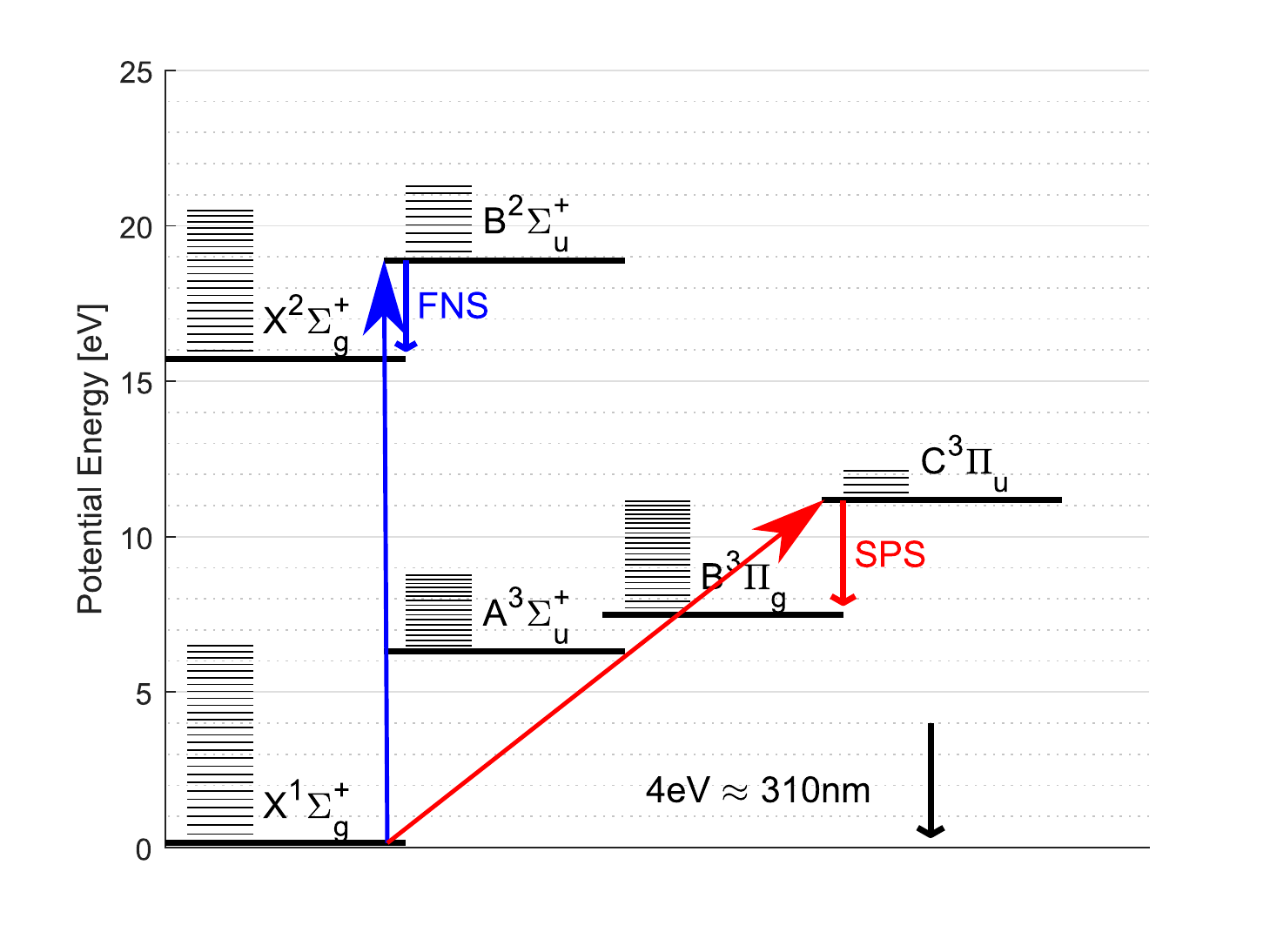}}}%
    \subfloat[ ]{
    \label{el_cross}
    {\includegraphics[width=0.5\textwidth]{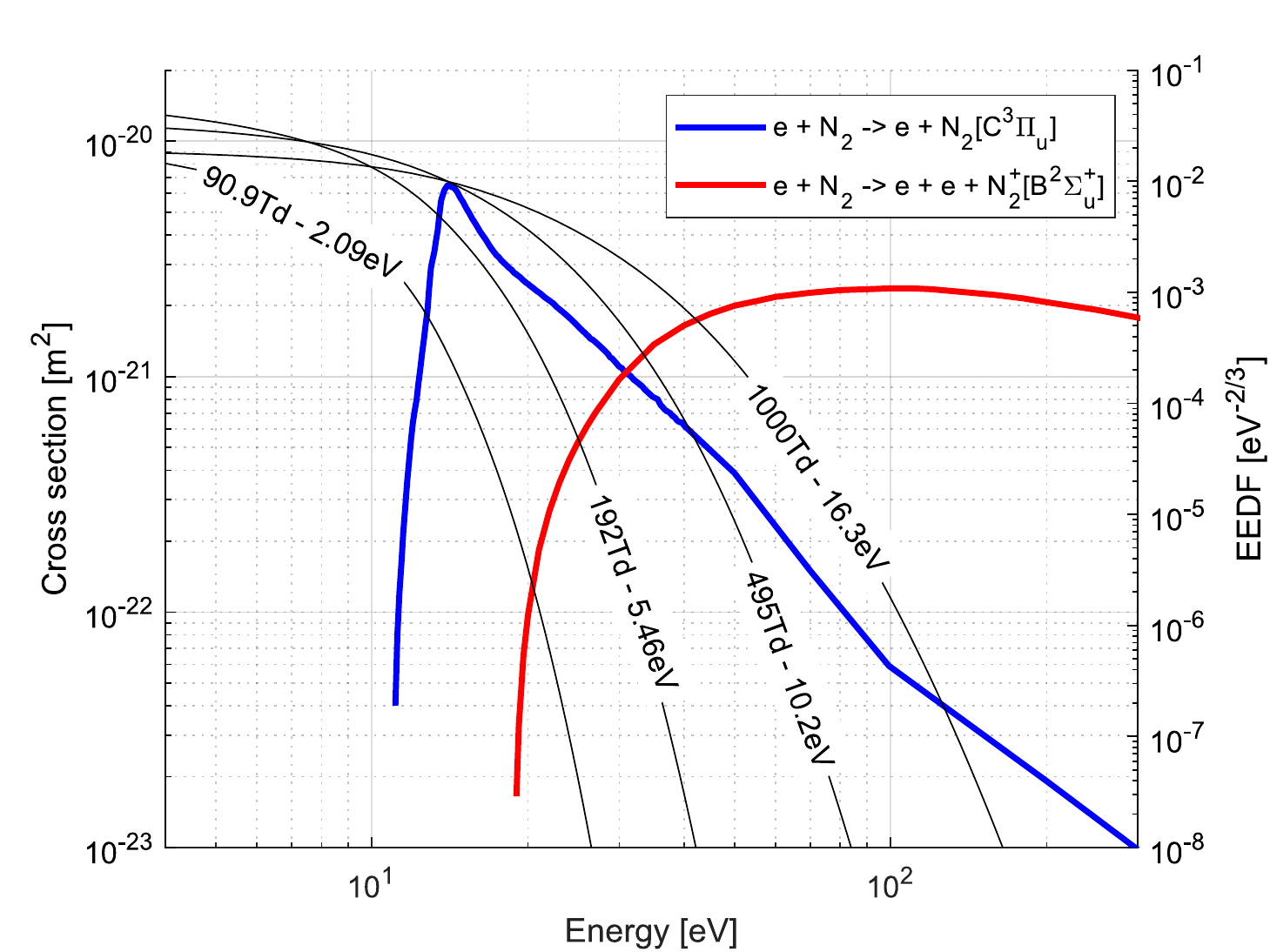} }}%
    \centering
    \caption{Fundamentals of the used intensity ratio method. a) Energy diagram~\cite{gaydon} for the most relevant electronic states of nitrogen with their vibrational levels, showing the pumping of the FNS and SPS upper states from the groundstate and their emissions. b) Integral electron-impact excitation cross section~\cite{Alves2014} for the ${\rm N_2}({\rm C}^3 \Pi_{\rm u})$ and ${\rm N^{+}_2}({\rm B}^2 \Sigma_{\rm u}^+)$ electronic states, combined with the EEDF's for various reduced electric field strengths in 33\,mbar pure nitrogen.}%
    \label{el_trans}%
\end{figure}

\begin{figure}
\centering
    \includegraphics[width=0.8\textwidth]{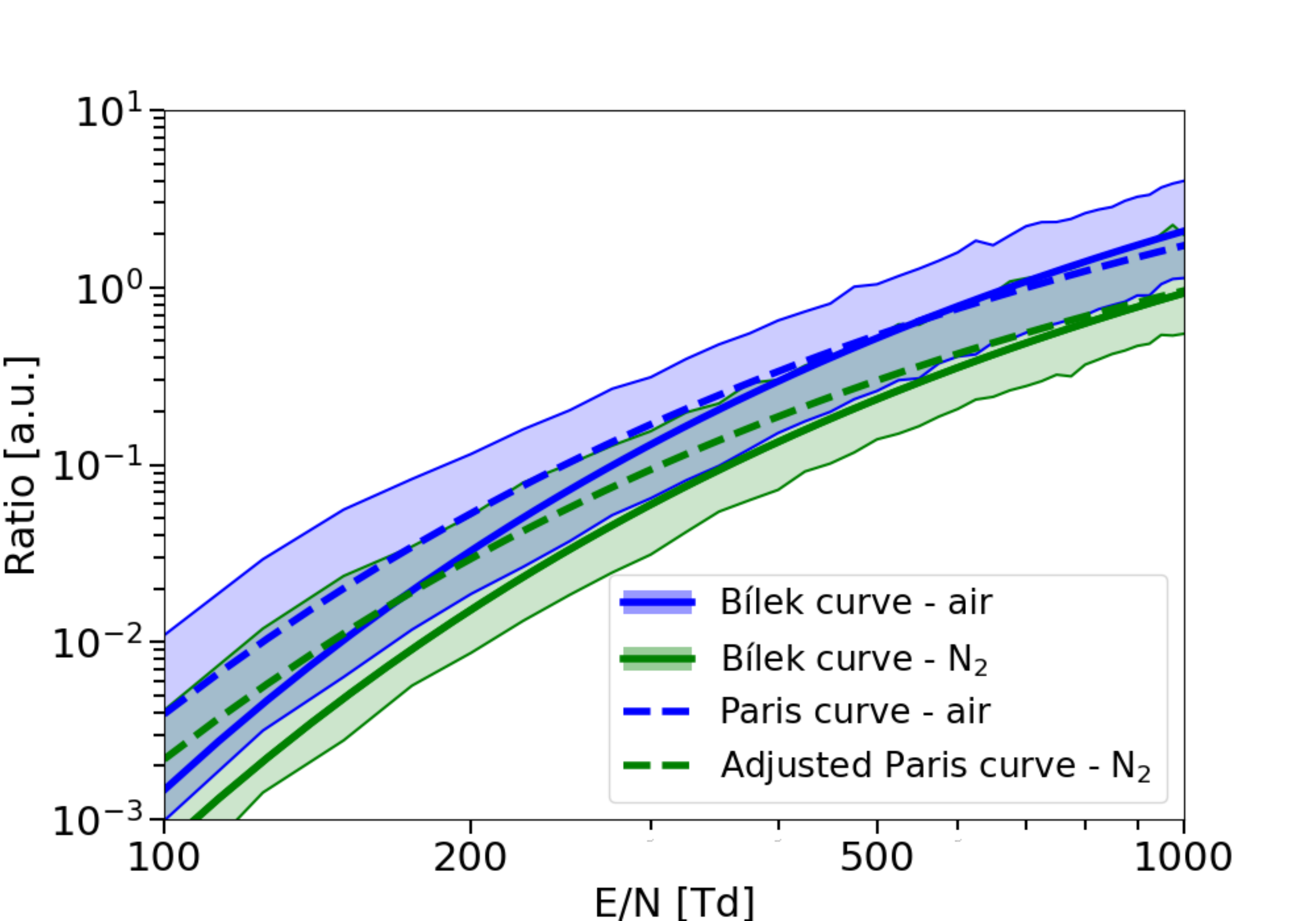}
\caption{The calibration curves, relating the ratio of FNS\,(0-0) and SPS\,(2-5) emissions to reduced electric field strengths, for Paris~\cite{paris2005intensity} divided by a factor of 2 and Bílek~\cite{bilek2018electric,bilek2019electric} for synthetic air and nitrogen at 33\,mbar.}\label{cal_curves}
\end{figure}

The electric field in nitrogen-containing plasmas can be determined from the intensity ratio of specific nitrogen spectral bands. 
Due to their radiative power, excitation energy and spectral positioning, the most used bands are the first negative system (FNS, ${\rm B}^2\Sigma_{\rm u}^+\rightarrow {\rm X}^2\Sigma_{\rm g}^+$) of N$_2^+$ and second positive system (SPS, ${\rm C}^3\Pi_{\rm u}\rightarrow {\rm B}^3\Pi_{\rm g}$) of N$_2$.
By far most of the emission radiated by streamer discharges comes from the SPS, making it the obvious first candidate for OES. 
To simplify the acquisition and calibration procedures, the dimmer vibrational transition SPS\,(2-5) was selected because the most intensive FNS transition, 0-0, is located spectrally in its vicinity, see also section~\ref{SpectralFitting}.
Figure~\ref{el_diag} shows the energy diagram of the most relevant nitrogen electronic states including their vibrational levels, with the transitions from and to the FNS and SPS upper levels indicated. 
The upper states of the FNS and SPS have different excitation thresholds for the population from the ground state, 11.3 and 18.7\,eV respectively.
In order to have a line ratio which is sensitive to the local electric fields in the discharge, their radiative upper states population mechanisms need to be susceptible to the EEDF.
Figure~\ref{el_cross} shows the electron impact cross sections for the population of the FNS and SPS upper states from the ground state.
In the same figure the EEDF's for various electric field strengths are depicted with their associated mean electron energy, these are calculated using a Bolsig+~\cite{Hagelaar2005} implementation in PLASIMO~\cite{Dijk2009} for 33\,mbar of nitrogen, see section~\ref{sec:model}.
The intensity ratio for these species is therefore valid for field strengths where both species are readily produced, but the mean electron energy of the EEDF is not so high that the sensitivity reduces, see the slope for the lines in figure~\ref{cal_curves}.

\setcounter{subfigure}{0}

The use of the intensity ratio method is limited by the fulfilment of the following conditions~\cite{bilek2019electric}.
Firstly the electron impact excitation of  N$_2^+$(${\rm B}^2\Sigma_{\rm u}^+$) and N$_2$(${\rm C}^3\Pi_{\rm u}$) states from level $\nu$=0 of the ground ${\rm N_2}({\mathrm X}^1 \Sigma_{\rm g}^+)$ state should be dominant. 
This requires limited gas heating and the majority of the nitrogen-ground-state molecules to be in the zeroth vibrational level while there is a limited density of meta-stable species.
Experimentally this means avoiding the spark transition since this is where most gas heating takes place.
To minimize the effect of meta-stables, a compromise is used for the repetition rate in our experiments.
A value is chosen such that there is enough signal accumulated within a reasonable time frame, the streamer is stable and has low inception jitter, but is still not too affected by meta-stables.
Secondly, the gas temperature should be around 300\,K.
The reaction rates are usually measured at 300\,K, when used at elevated temperatures, temperature dependencies of these rates need to be determined.

When the previous conditions are satisfied, the dependence $R(E/N)$ relating the intensity ratio and the reduced electric field is given by~\cite{kozlov2001spatio,Goldberg2022}:
{\def\d{{\rm d}}
\begin{equation}
	{R}\left(E/N\right) = \frac{p^{\rm \scriptscriptstyle FNS}_{ \nu'\!\nu''} k^{\rm \scriptscriptstyle B}_{\rm \scriptscriptstyle im}(E/N) \tau_{\rm eff}^{\rm \scriptscriptstyle FNS}}{p^{\rm \scriptscriptstyle SPS}_{ \nu'\!\nu''} k^{\rm \scriptscriptstyle C}_{\rm \scriptscriptstyle im}(E/N) \tau_{\rm eff}^{\rm \scriptscriptstyle SPS}} = \frac{\tau_{\rm eff}^{\rm \scriptscriptstyle FNS}}{\tau_{\rm eff}^{\rm \scriptscriptstyle SPS}}\frac{\displaystyle \strut \frac{\d I_{\rm \scriptscriptstyle FNS}}{\d t} + \frac{I_{\rm \scriptscriptstyle FNS}}{\tau_{\rm eff}^{\rm \scriptscriptstyle FNS}}}
	{\displaystyle \strut \frac{\d I_{\rm\scriptscriptstyle SPS}}{\d t} + \frac{I_{\rm\scriptscriptstyle SPS}}{\tau_{\rm eff}^{\rm \scriptscriptstyle SPS}}},
\label{full-time-eq}
\end{equation}}where
$k^{\rm \scriptscriptstyle B, \rm \scriptscriptstyle C}_{\rm \scriptscriptstyle im}$ are electron-impact rate constants for population of  N$_2^+$(${\rm B}^2\Sigma_{\rm u}^+$) and N$_2$(${\rm C}^3\Pi_{\rm u}$) states. 
$p^{\rm \scriptscriptstyle FNS, \rm \scriptscriptstyle SPS}_{ \nu^\prime\!\nu^{\prime\prime}}$ is the emission line radiative power:
\begin{equation} 
	p_{ \nu^{\prime}\!\nu^{\prime\prime}} = \frac{ hc}{ \lambda_{\nu^{\prime}\!\nu^{\prime\prime}}}\frac{1}{\tau_{\rm \scriptscriptstyle 0}^i} , \qquad {\rm [W]}
  \label{species_density}
\end{equation}
where $\tau_{\rm 0}^i$[s] is the lifetime of spontaneous emission, $i$ defines the excited state, in our case $i \in \{{\rm FNS, SPS}$\}, and $\lambda_{\nu^{\prime}\!\nu^{\prime\prime}}$ is the characteristic wavelength of emitted radiation during the transition between the upper ($\nu^{\prime}${}) and the lower ($\nu^{\prime\prime}${}) vibronic states.
The $I_{i}$ are the measured intensities of respective spectral bands.
The effective lifetimes $\tau_{\rm eff}^{ \scriptscriptstyle i}$ of the excited state in synthetic air and nitrogen can be expressed as
\begin{equation}
\frac{1}{\tau_{\rm eff}^i} = \frac{1}{\tau_{\rm \scriptscriptstyle 0}^i} + N \left(f_{\rm N_2} k^{i}_{q, \rm N_2} + f_{\rm O_2} k^{i}_{q, \rm O_2}\right),
\label{eff_life}
\end{equation}
with $f$ the fraction of gas components: $f_{\rm N_2} = 0.8$, $f_{\rm O_2} = 0.2$ for synthetic air and $f_{\rm N_2} = 1.0$, $f_{\rm O_2} = 0.0$ for pure nitrogen.
$N$ defines the volumetric gas particle density of the mixture.

The ratio ${R}\left(E/N\right)$ can be obtained by either using a Townsend discharge setup~\cite{paris2005intensity} or by using a kinetic model with appropriate kinetic data~\cite{bilek2019electric}. 
Within this work, we use the intensity ratio of the FNS\,(0-0) and SPS\,(2-5) to determine the electric field in synthetic air and pure nitrogen at 33\,mbar. 
A calibration curve for FNS\,(0-0)/SPS\,(2-5) in synthetic air at atmospheric pressure was determined experimentally by Paris~\cite{paris2005intensity} and has the following form:
\begin{equation}
R_{\rm \scriptscriptstyle FNS\,(0-0)/SPS\,(2-5)}(E/N) = a \exp{\left[-b(E/N)^{-0.5}\right]} \qquad \text{with} \qquad a = 46.0, b = 89.0.
\label{paris_curve}
\end{equation}
According to recent research~\cite{hoder2016radially}, the parameter $a$ of this equation should be divided by a factor of two. 
For use in reduced pressure experiments the ${R}\left(E/N\right)$ is extended with the effective lifetimes as follows:
\begin{equation}
	{R}\left(E/N, N \right) = \frac{\tau_{\rm eff}^{\rm \scriptscriptstyle FNS}(N)}{\tau_{\rm eff}^{\rm \scriptscriptstyle FNS}(N_0)} \frac{\tau_{\rm eff}^{\rm \scriptscriptstyle SPS}(N_{\rm \scriptscriptstyle 0})}{\tau_{\rm eff}^{\rm \scriptscriptstyle SPS}(N)}{R}\left(E/N, N_{\rm \scriptscriptstyle 0} \right),
\label{scaling-pressure}
\end{equation}
where $N_{\rm \scriptscriptstyle 0}$ is the gas density at atmospheric pressure.

The calibration curve can be calculated from a simple kinetic model based on the most important processes as listed in Table~\ref{IMprocesses}. 
The processes R1 and R2 are electron-impact excitations, R3 and R4 spontaneous emissions and R5-R8 quenching reactions depopulating the upper states by collisions with N$_2$ and O$_2$.
The calibration curves, shown in figure~\ref{cal_curves}, were calculated using the reaction rate constants as defined in the third column of table~\ref{UQ:intervals}. 
These reaction rates were used to calculate the calibration curves as presented in equation~\eqref{full-time-eq}.
In order to establish the uncertainty of the electric field determination, we define uncertainties of the reaction rates used in the kinetic model, see the fourth column of table~\ref{UQ:intervals}. 
The defined uncertainty ranges are discussed in detail in~\cite{bilek2019electric, obrusnik2018electric,bilek2018electric}. 
The final uncertainty calculation of the calibration curves is now performed by Monte-Carlo-based uncertainty quantiﬁcation, where rate constants were varied within the range as defined in the fourth column of table~\ref{UQ:intervals}. 
The uncertainty of the electron impact rate coefﬁcients (R1, R2) was obtained by combining the electron-impact cross-sections (denoted CS in table~\ref{UQ:intervals}) of R1 and R2 with different electron-energy distribution functions (EEDFs) obtained from the cross-section libraries (Biagi~\cite{Biagi}, Phelps~\cite{Phelps}, Triniti~\cite{TRINITI}, IST Lisbon~\cite{Alves2014}). 
This approach is described in detail by Obrusnik et al.~\cite{obrusnik2018electric}. 

The calibration curves of Paris divided by a factor of 2 and the calibration curves obtained by the kinetic model from the previous paragraph, can be described by equation~\eqref{paris_curve} for 33\,mbar. 
The parameters of this equation are listed in table~\ref{bounds} and are plotted in figure~\ref{cal_curves}. The Paris calibration curve including the correction factor of 2 for synthetic air is adjusted to use pure nitrogen instead of air, using identical rate constants for electron-impact excitation.  

The synthetic air calibration curves lie above the pure nitrogen calibration curves, see figure~\ref{cal_curves}. This is due to the quenching of the N$_2$(${\rm C}^3\Pi_{\rm u}$, $\nu$=2) state by O$_2$, which is about 8 times higher in air than in pure nitrogen. 
This change reduces the effective lifetime of the N$_2$(${\rm C}^3\Pi_{\rm u}$) state, which is in the denominator of equation~\eqref{full-time-eq}, and therefore shifts the calibration curve of air upwards compared to pure nitrogen. 
This leads to the important implication that the FNS\,(0-0)/SPS\,(2-5) intensity ratio has a higher value in air than in pure nitrogen at the same reduced electric field, which is also visible in figure~\ref{spectral_fits_exp} where the peak field is plotted for synthetic air and pure nitrogen with otherwise identical conditions. 

\subsection{Spectral fitting}\label{SpectralFitting}
\begin{figure}[h!]
    \centering
 \includegraphics[width=1\textwidth]{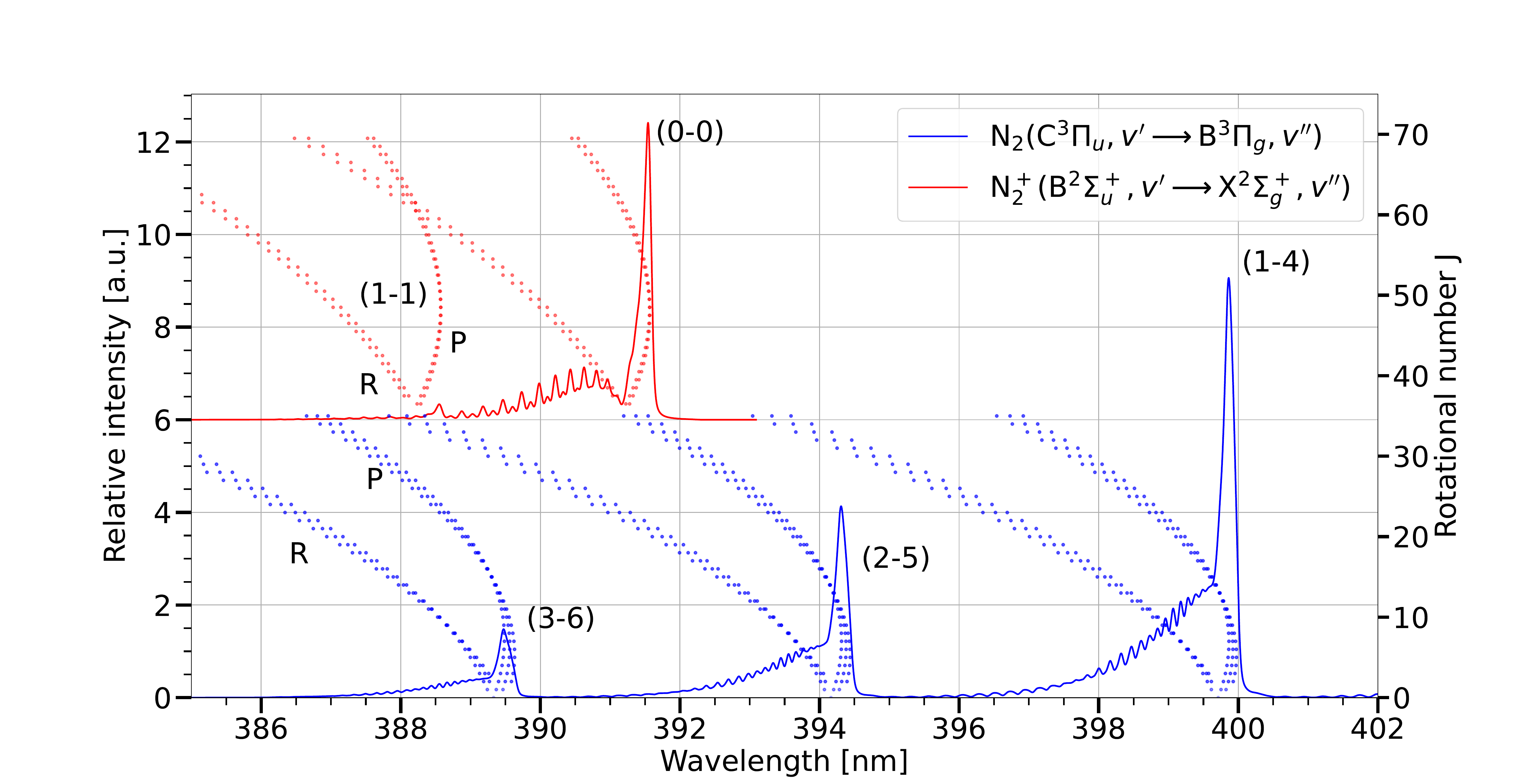}
\caption{Separated simulation of FNS and SPS vibrational bands with increased spectral resolution in comparison to our typical measurements. For illustrative purposes, the individual major rotational branches are displayed as dots. For clarity, the relative intensity and rotational number of FNS have been shifted up by 6 and 35, respectively. Where the numbers in brackets represent higher ($\nu^{\prime}$) and lower ($\nu^{\prime\prime}$) vibrational states, respectively. \textit{MassiveOES} databases used for the simulation are based on \cite{luque1999lifbase,Nassar_2004,LAUX19929,Faure_1998}.}
\label{seperation_overview}
\end{figure}

\begin{figure}%
    \centering
    \subfloat[]{{\includegraphics[width=0.45\textwidth]{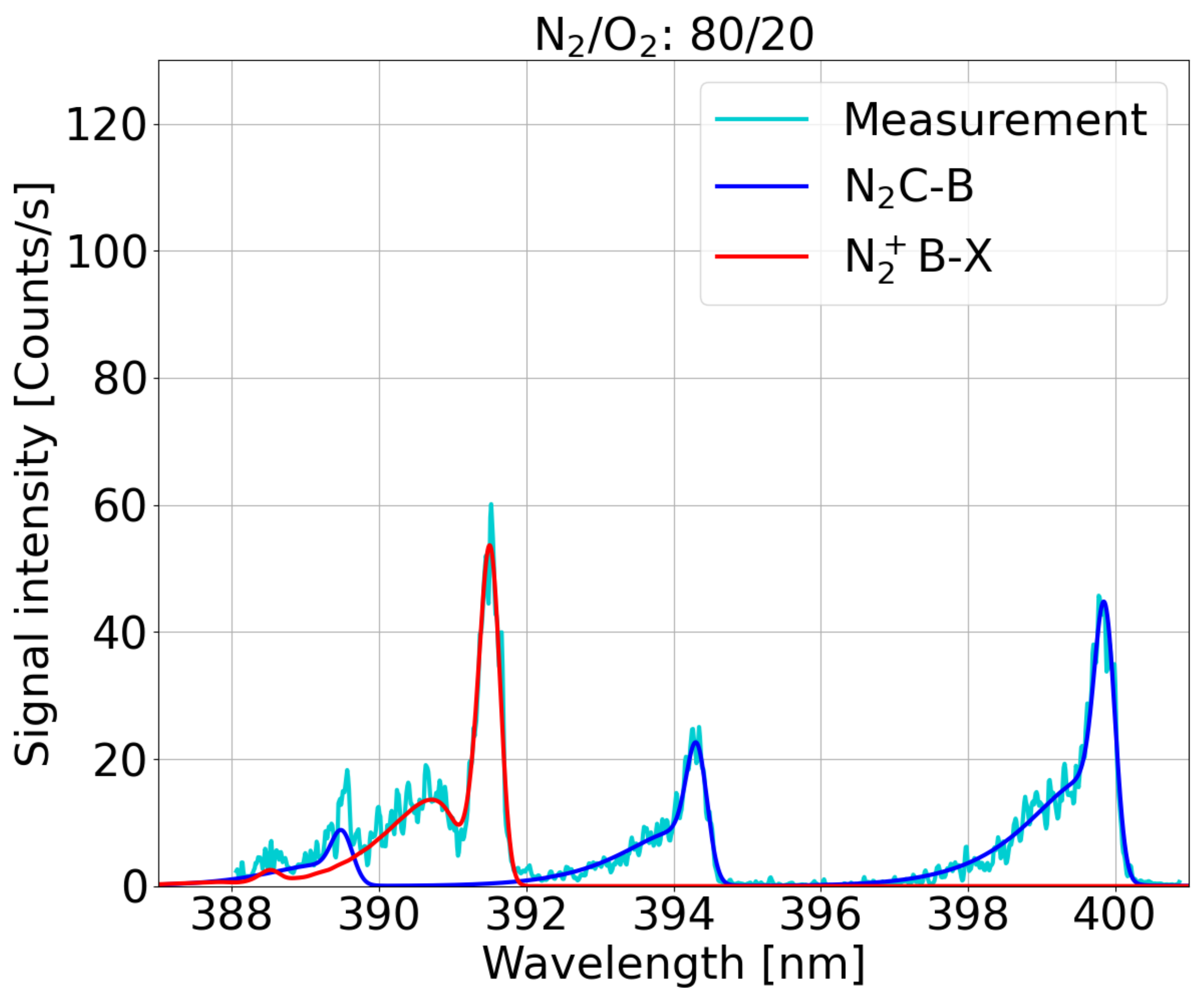} }}%
    \qquad
    \subfloat[]{{\includegraphics[width=0.45\textwidth]{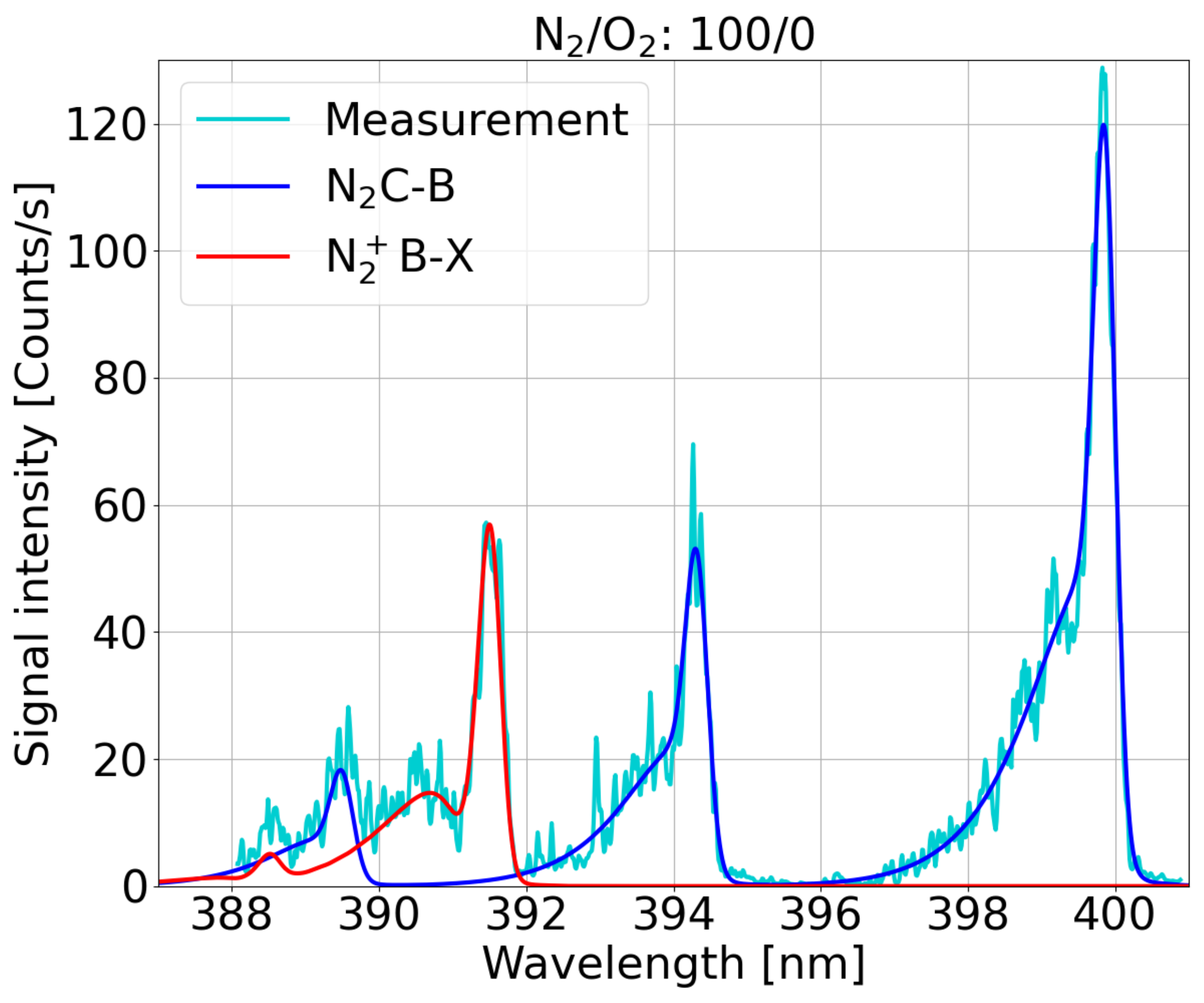} }}%
    \caption{Spectral fits of a) synthetic air and b) pure nitrogen at 33\,mbar and 8\,kV, 500\,ps gated spectra, taken at 7\,ns and 10\,ns, respectively, see figure~\ref{33mbarAxial}. The fit of FNS and SPS is separated to demonstrate possible overlaps of FNS\,(0-0), SPS\,(2-5) and SPS\,(3-6). It should be noted, that intensities are divided by the accumulation time of the ICCD camera to illustrate the intensity/ratio difference between pure nitrogen and synthetic air.}%
    \label{spectral_fits_exp}%
\end{figure}

The spectral `fingerprint' of the propagating space charge layer, is depicted in figure~\ref{spectral_fits_exp}.
The spectrum, centered around 394\,nm with a window of 14\,nm shows features of both the SPS and FNS.
The raw spectral images from the ICCD are firstly wavelength calibrated using an HgAr source, additionally the raw 2D images are shear-corrected for symmetry.
Secondly, the intensity of the 2D image is corrected using a broadband white light source.
Lastly, some light pixel noise filtering and smoothing is applied.
Additional smoothing is applied for the spectra where an Abel inversion is performed.
The Abel inversions themselves are performed using an implementation of Dribinski~\cite{Dribinski2002} in PyAbel~\cite{Gibson2022}.
The specific states used for the calculation of the ratio method are the SPS (2-5) and FNS (0-0):
\begin{gather*}
\mathrm{N}_2(\mathrm{C}^3\Pi_{u})_{\nu^{\prime}=2}\rightarrow \mathrm{N}_2(\mathrm{B}^3\Pi_{g})_{\nu^{\prime\prime}=5}+h\nu, \quad\lambda=394.3 \mathrm{\,nm} \\
\mathrm{N}_2^+(\mathrm{B}^2\Sigma_{u}^+)_{\nu^{\prime}=0}\rightarrow\mathrm{N}_2^+(\mathrm{X}^2\Sigma_{g}^+)_{\nu^{\prime\prime}=0}+h\nu,  \quad\lambda=391.4 \mathrm{\,nm}
\label{FNS_SPS_Transitions}
\end{gather*}
where $\nu^{\prime}$ is a vibrational upper state and $\nu^{\prime\prime}$ is a vibrational lower state.
Since both species are captured together, no additional calibration needs to be performed.
All features in the aforementioned measured spectral window are fitted, figure~\ref{seperation_overview} shows the individual contributions to the visible spectrum.
The figure illustrates the simulation of our measured spectral range with separated FNS and SPS components. 
The simulation has increased spectral resolution for illustrative purposes and includes individual rotational states for every vibrational transition, it should be noted that only the major rotational branches (R: $\Delta J \mathrm{\,}$=$ \mathrm{\,}$1 and P: $\Delta J \mathrm{\,}$=$ \mathrm{\,}$-1) are displayed for the sake of clarity.
The rotational branches demonstrate the density of rotational states in N$_2$ and suggest the reason why optical diagnostics of nitrogen is generally challenging and requires substantial spectral resolution together with a comprehensive evaluation approach. 

For the determination of rovibrational temperatures and the intensities of the SPS and FNS bands with the high spatio-temporal resolution, a routine based on $\textit{massiveOES}$ software was deployed~\cite{massiveoes}. 
The used fitting routine assumes thermal equilibrium (Boltzmann distribution) of rotational and vibrational quantum states and allows batch processing of spectra, achieving a time-efficient calculation.
For an experiment consisting of 100 radial bins and 25 temporal steps, 2500 spectra are fitted.
The rotational and vibrational temperatures are determined from vibrational bands SPS\,(1-4),~(2-5),~(3-6) and FNS\,(0-0),~(1-1) and their entire rotational manifold.

It should be emphasized that the thermal fitting used for the determination of rovibrational temperatures by definition assumes a Boltzmann distribution, that is the distribution of rotational and vibrational states can be described by using single rotational and vibrational temperatures. 
In cases where the distribution is non-thermal, meaning that the distribution cannot be described by a single temperature, more sophisticated approaches are required. 
In the case of partial thermalization, it is often favourable to use a two or higher-term Boltzmann distribution (assuming two or three temperatures), see for example \cite{Wang_2007,Bruggeman_2009} or take a more general state-by-state approach.

This latter approach considers individual quantum states as separate species and as such the temperature as a parameter that defines the ratio of the populations of individual states loses its meaning.
On the other hand, the state-by-state approach allows us to determine the populations of individual states separately and investigate the distribution of populations without any assumption of a thermalized system. 
Based on the distribution of the resulting populations it is then relatively easy to determine if the investigated system is thermalized, partially thermalized or has some local overpopulation of specific states.
This approach is not only time-efficient, but also the precision of population determination is enhanced in comparison to traditional methods. 
For more information about this approach see the work of Vor\'{a}\v{c} et al.~\cite{massiveoes,vorac2017}.
The semi-automated state-by-state temperature-independent fitting of molecular spectra is also part of the $\textit{massiveOES}$ software.
The fitting routine based on the state-by-state approach was used to investigate the equilibrium of rovibrational quantum states presented in figure~\ref{seperation_overview}.


The intensities used for the intensity ratio calculation of the reduced electric field are determined from the fitted spectra.
In figure~\ref{spectral_fits_exp} examples of fitted spectra are shown for synthetic air and pure nitrogen.
The figure illustrates the effect of quenching of SPS states by O$_2$, while FNS states are seemingly unaffected.

By fitting the spectra and using the total intensities of all rotational branches belonging to a specific transition, accurate relative intensities are determined for the intensity ratio. 
An alternative option is to determine the peak values of band heads as intensities.
Nevertheless, this approach can lead to several issues.
Firstly, the detected signal can differ significantly for different stages of the investigated discharge, as the rotational structure of molecular spectra is not necessarily the same for every stage~\cite{hoder_2016_loft}. 
Secondly, the intensity of a band head is often not proportional to the total vibrational transition radiation intensity. 
Instead, the use of the surface under the measured vibrational band results in more precise values as it includes all of the rotational states. 
Therefore, the intensities used in equation~\eqref{full-time-eq} are determined as the surfaces of the SPS\,(2-5) and FNS\,(0-0) vibrational transitions. 
Another benefit of using the surfaces of the simulations is the suppression of the overlap in the measured FNS and SPS spectra, which can cause a small error in the determined ratio, especially if the higher rotational states are dominant. 
This effect can be seen in figure~\ref{seperation_overview}, where even states with the rotational number $\textrm{J}=20$ of SPS\,(2-5) are beginning to influence the total intensity of FNS\,(0-0).
More information and more detailed comparison regarding the use of peak values or integral values can be found in \cite{bilek2019electric} together with a discussion regarding the adjustment of equation~\eqref{full-time-eq} for the use of peak values for the determination of intensities.

A different experimental approach for determining the reduced electric field from the intensity ratio method is by using photo multipliers instead of an ICCD.
Time correlated single photon counting~\cite{kozlov2001spatio,Hoder_Cernak,kusyn} has the advantage of a much improved temporal resolution, going down to tens of picoseconds, compared to the hundredths of picoseconds generally attained by an ICCD, this matters especially for the time derivatives in the ratio calculation.
However, interpretation-wise, the ICCD provides directly correlated spatial and spectral dimensions to the measurement, allowing the spectral fitting of the captured data and spatial verification of the 2D single-shot imaging.

\setcounter{subfigure}{0}

\section{Results and discussion}
\begin{figure}%
    \centering
        \subfloat[synthetic air]{
        \label{33mbarAirAxial}%
        {\includegraphics[width=0.5\textwidth]{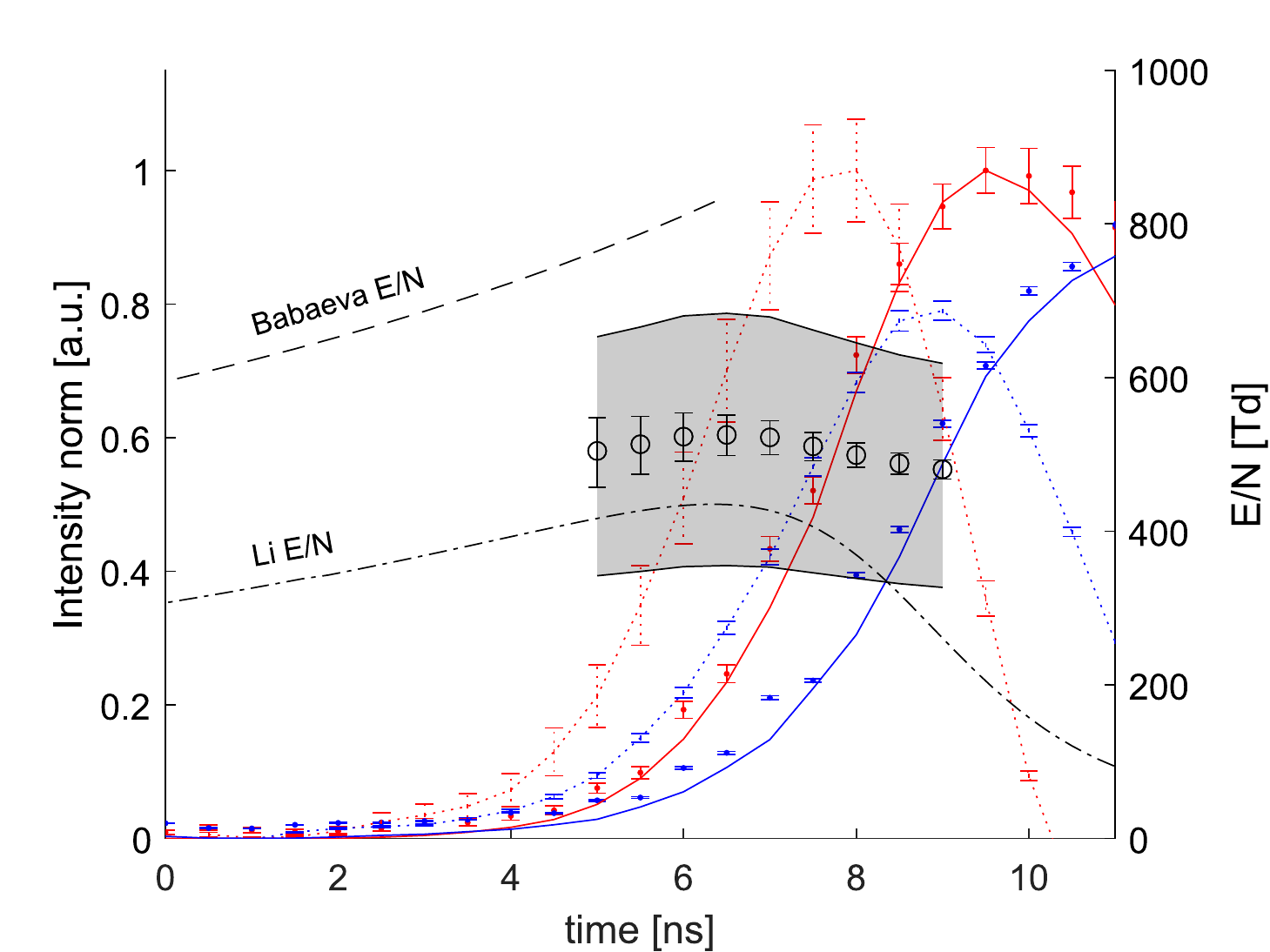} }

    }%
    \subfloat[pure nitrogen]{
        \label{33mbarN2Axial}%
        {\includegraphics[width=0.5\textwidth]{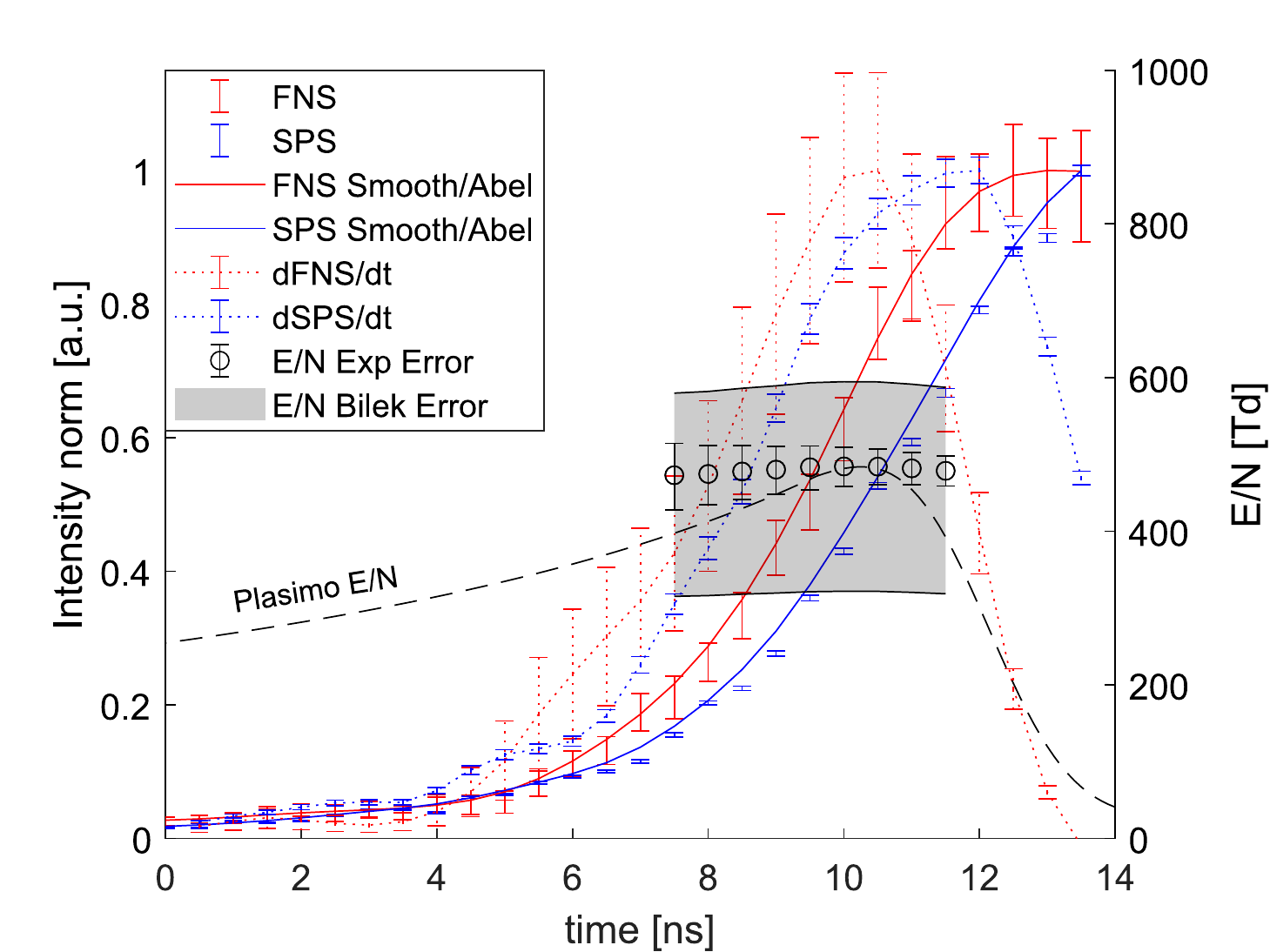} }
    }%
    \caption{FNS and SPS signals with their derivatives, derived reduced electric fields and simulation results for comparison. For synthetic air comparisons to theoretical literature (Babaeva and Naidis~\cite{Babaeva2021}) and numerical modelling (Li et al~\cite{Li2021}) are included, while for pure nitrogen the $E/N$ curve used as input of the global model is indicated. Red and blue solid lines plot smoothed data for pure nitrogen and Abel inverted data for synthetic air, otherwise the legend is shared between the figures.}%
    \label{33mbarAxial}%
\end{figure}

\begin{figure}
    \centering
 \includegraphics[width=1\textwidth]{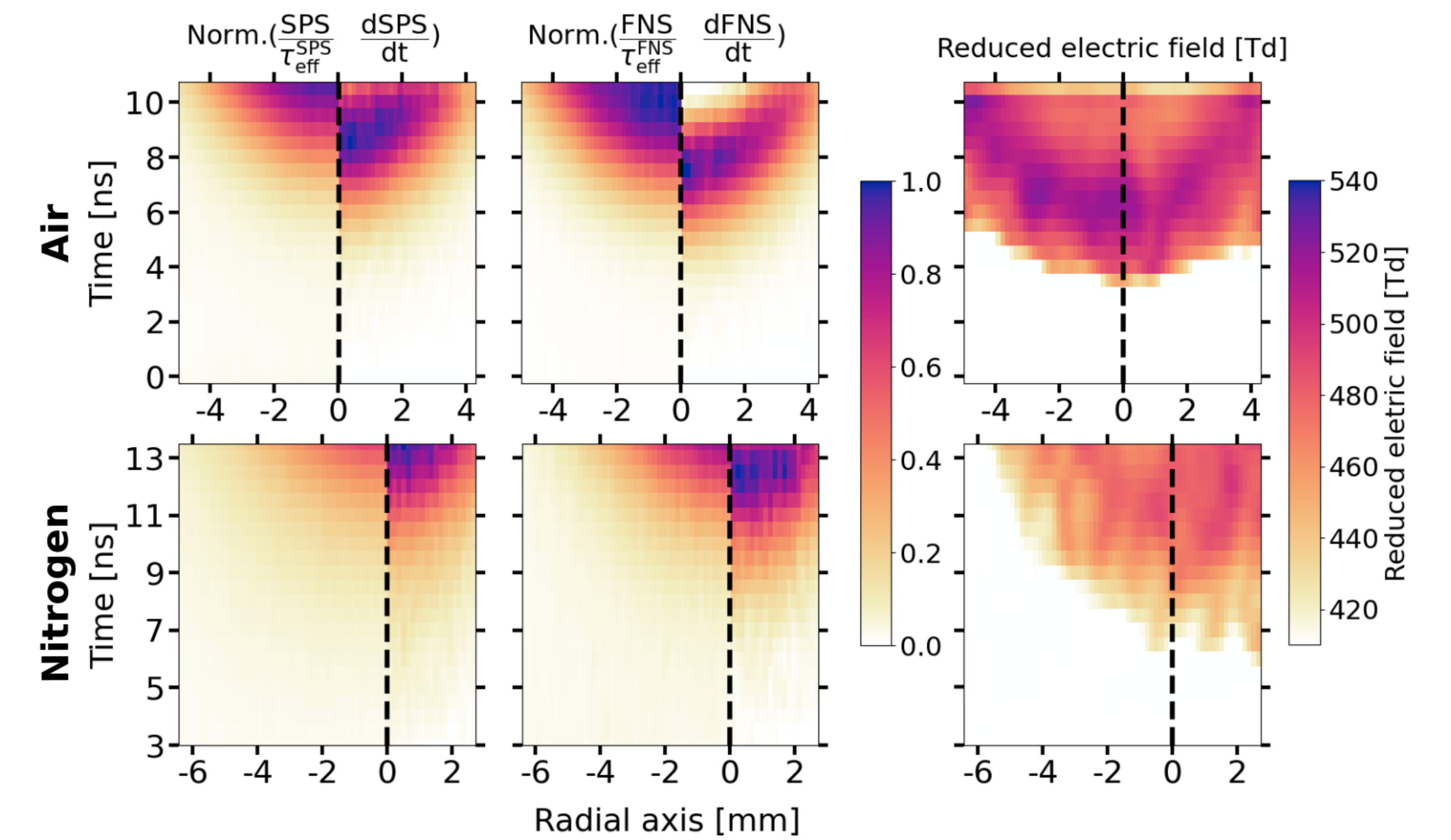}
\caption{Heat maps of downward facing streamers, showing the spatial and temporal profiles of the variables used in calculating the reduced electric field and of the resulting reduced electric field.
Abel inversion was not applied to the data displayed in this figure.
Only half of the data is shown due to the assumed symmetry, the dashed line indicates this symmetry axis. For the reduced electric field the full spatial data range is displayed, while the field is cut off at 420\,Td.
Intensities are divided by their respective effective lifetimes (left side) and their derivatives (right side) for SPS\,(2-5) and FNS\,(0-0). The upper row is for synthetic air and the lower row is for pure nitrogen. All of the presented data underwent Gaussian smoothing.}
\label{fields_heatmap}
\end{figure}

In figure~\ref{33mbarAxial} the intensity ratio variables are presented for synthetic air and pure nitrogen.
Figure~\ref{fields_heatmap} contains heat maps of intensity ratio variables including the radial axis, from which the spatially central data bin is shown in figure~\ref{33mbarAxial}.
The figure shows the intensity of streamer emissions accumulated as line integrals at sequential time delays.
The smoothed data and its time derivative are plotted on opposite sides of a symmetry axis in the same figure, as the discharge is generally symmetric over the axial axis.
The time derivatives displayed on the right-hand side of the figure are calculated from the right-hand side of the data.
The data for the pure nitrogen experiment is radially slightly shifted off-axis, since the streamer path for this experiment was slightly off-center.
This is also visible in figure~\ref{ratios_streamers}, where the pure nitrogen streamer shows some lateral movement near the centre of the gap before it straightens when nearing the grounded electrode.
The intensities and their derivatives show the hemispherical nature, see the curved signal distributions in figure~\ref{fields_heatmap}, of the imaged space charge layer. 
The intensities and their derivatives underwent Gaussian smoothing with sigma parameters of 0.75\,ns and 210\,$\mu$m in temporal and spatial (radial) dimensions. 
The resulting reduced electric field determined from smoothed intensities underwent further Gaussian smoothing with a parameter of 210\,$\mu$m for the x-axis (radial component) and 0.25\,ns for the y-axis (temporal component).

Because the accumulated signals are line integrals, the discharge is cylindrically symmetric and we are interested in the radial distribution, an Abel inversion is applied to the fitted $I_\text{FNS}$ and $I_\text{SPS}$.
In figure~\ref{33mbarAirAxial} the Abel inverted data (solid line) plotted alongside the smoothed non Abel inverted data (markers with error bars) shows that $I_\text{FNS}$ decreases more strongly after its peak.
This is due to the line integral accumulation of $I_\text{FNS}$ emission from the hemispherical streamer head and thus significant emissions from around the equator of the hemisphere are accumulated into the data.
This means that behind this peak the calculated reduced field strengths are overestimated when no Abel inversion is applied while for earlier time steps, the Abel inversion has less effect due to the small width of the discharge.
The calculations of derivatives and $E/N$ are performed on the non Abel inverted data due to the amount of smoothing needed to apply a proper Abel inversion.

The calculated reduced electric field is only shown for a restricted window.
For earlier time steps the low total intensity causes a large error in the derivatives, resulting in inaccurate calculations.
The calculated data is only shown for the region where enough signal is accumulated.
The upper limit is set at the maximum of the $I_\text{FNS}$ derivative, since this point in time coincides roughly with the position of the space charge layer, after which the local electric field strongly decreases and an Abel inversion should be applied, see also figure~\ref{Plasimo_a} and Shcherbakov et al~\cite{shcherbakov2007}.
The calculated reduced electric field data for the non Abel inverted data is therefore cut off at this limit.
The error bar markers show the error in fitting the spectra, while the error region, marked as the shaded area between the black lines, depicts the error margin in the calculation of the reduced electric field due to the uncertainty of reaction rates and lifetimes as discussed in section~\ref{SpectralFitting} and~\cite{bilek2019electric}.

As a comparison, firstly the field calculated with the methods by Babaeva and Naidis~\cite{Babaeva2021} is plotted in figure~\ref{33mbarAirAxial}.
There, a peak field $E/\delta$ of 200\,kV/cm for atmospheric pressure air is mentioned, where $\delta$ is the pressure fraction.
This peak field corresponds with the field directly adjacent to the space charge layer, in order to plot the field as a function of time for the central axis plotted in figure~\ref{33mbarAxial}, the experimentally determined velocity and diameter of the streamer are used.
This is because the electric field distribution is affected by the local field enhancement and thus the diameter of the streamer and the velocity of the streamer affects the elongation of this curve. 
For the 33\,mbar streamer in synthetic air a $0.5\times1.7$\,cm (0.5 as the ratio between the optical diameter and the high field diameter~\cite{Babaeva2021}) thick streamer propagating with $5.3\cdot 10^5$\,m/s is used derive the time dependency.
Secondly, the simulated electric field profile of a streamer based on modelling from Li et al~\cite{Li2021} is plotted.
The conditions of Li's work are adapted to the used experimental setup and for this comparison, the simulation was re-run for the voltage and pressure used in this experiment.
The calibration curve data for Paris mentioned in~\ref{cal_curves} give nearly identical results for the ratios, and thus fields, around 500\,Td, since this is where both calibration curves intersect.

For the results in pure nitrogen, shown in figure~\ref{33mbarN2Axial}, a direct comparison with model results was unfortunately not yet available, due to issues with simulating cylindrically symmetric streamers without a dominant free electron source like photoionization.
However, Kulikovsky~\cite{Kulikovsky1995} also mentions a peak electric field strength of 200\,kV/cm for a streamer in  atmospheric pressure nitrogen, which is identical to the value given by Babaeva et al~\cite{Babaeva2021} for air.

\setcounter{subfigure}{0}

The peak reduced electric fields derived from the measurements for synthetic air and pure nitrogen at 33\,mbar are approximately 540 and 480\,Td respectively, which is very close, especially considering the uncertainties in both the experiment and the calculation.
Especially the uncertainty in the calibration curve, due to the scatter of literature data on cross-sections for nitrogen and air kinetics is large.
While the experiments are performed using identical conditions except for the used gas mixture, there are observable changes in the streamer due to the admixture of oxygen.
The main observable difference is the optical diameter of the streamers, 1.7 and 1.1\,cm at 33\,mbar for synthetic air and pure nitrogen respectively, which would suggest more field amplification and thus a higher peak field for pure nitrogen.
Additionally, the higher efficiency of photoionization in the synthetic air mixture lowers the field needed for streamer propagation~\cite{Nijdam2010}.
In synthetic air free electrons are created further ahead of the space charge layer, and those can avalanche towards the streamer, while in pure nitrogen nearly all ionization takes place closer to the space charge layer.
The ionization coefficient for air is slightly higher than for pure nitrogen, i.e., it can also ionize (and thus avalanche) more efficiently than pure nitrogen~\cite{Raizer2011}.
All these differences would suggest a higher peak $E/N$ for pure nitrogen, which is not what we observe.
This could be due to the assumption of completely smooth space charge layers, which might not be the case for pure nitrogen~\cite{Nijdam2010} at the pressure and level of oxygen contamination in the experiment.
In other words, the line integrated spectra of fitted FNS and SPS species might originate from multiple local electric field maxima, over the more stochastic or irregular looking hemispherical space charge layer, see Bagheri et al~\cite{Bagheri2019}.

\begin{figure}
\centering
    \includegraphics[width=1\textwidth]{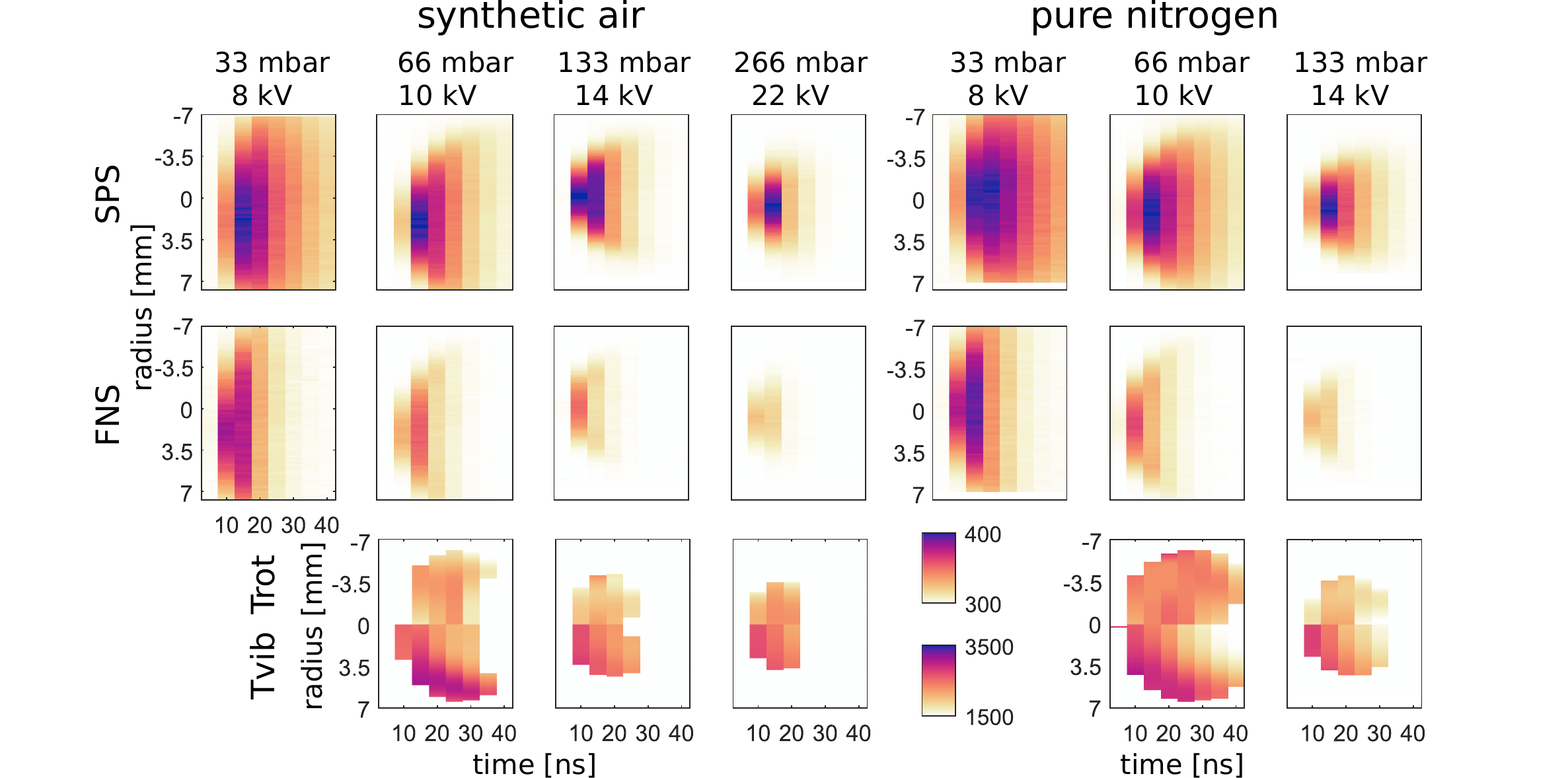}
\caption{Array of heat maps, streamers are shown propagating to the left, showing the intensities of SPS and FNS as well as the rotational and vibrational temperatures of SPS for various pressures in synthetic air and pure nitrogen. Spectra are captured with a 5\,ns temporal gate and step. FNS intensities are normalized to the SPS intensities. The bottom row shows temperatures fitted to the Abel inverted spectra.  }\label{RadialHeatmaps}
\end{figure}

Figure~\ref{RadialHeatmaps} contains an array of images representing the fitted parameters of the spectral space charge layer.
Each column represents a different discharge condition, the first four show different pressures and voltages in synthetic air, the last three in pure nitrogen.
The experimental pressures range from 33 to 266\,mbar, doubling between experiments.
The voltages are set to the minimal value for which each experiment has a stable crossing streamer.
At pressures below 30\,mbar, the streamer diameter became so large compared to the gap distance that it can no longer be regarded as a free moving streamer in bulk gas.
At pressures above 400\,mbar, in the used electrode geometry, streamer discharges always branch, which is already the case in 266\,mbar pure nitrogen, which is why these are omitted from this overview.
The first two rows show the fitted $I_\text{FNS}$ and $I_\text{SPS}$ values measured in the middle of the gap as a function of time and radial position.
In the bottom row, the fitted rotational and vibrational temperatures from Abel inverted data of the SPS are shown.
Contrary to the Abel inversion applied to the 500\,ps measurements discussed above and shown in figure~\ref{33mbarAxial}, first the Abel inversion is applied to the spectrum and then the inverted spectrum is fitted.
Each spectral bin of approximately 0.02\,nm is individually Abel inverted, to create the radially resolved heat maps.

What is clearly visible from these heat maps is the shrinking diameter of the streamer when increasing the operating pressure.
While the voltage is increased along with the pressure, it is not increased proportionally to maintain a constant $V/p$ (applied voltage over pressure) and therefore background $E/N$.
However, while the streamer diameter decreases, the local field amplification will compensate for this in a way that $E/N$ at the streamer head will be roughly constant across pressures, in accordance with the similarity laws stated by Briels et al.~\cite{Briels2008}.
It should be noted that in Briels et al. the minimum diameter of the streamer is defined.
The streamers created for our experiments are tuned for repetitiveness, always crossing the gap and not branching.
This means their diameter can be shrunk by decreasing the applied voltage at a certain pressure, but this causes the streamer to not necessarily cross the gap, inducing stochastic effects. 

In addition to the parameter variations discussed above, we also varied the repetition rate of the discharge from 20 to 200 and 1000\,Hz, but only at 1000\,Hz significant additional heating was measured: a few Kelvin between 20 and 200\,Hz and up to around 10\,K for 1000\,Hz.
For this research, it was decided to stay well below the 1\,kHz.

\subsection{Comparison to Global model}\label{sec:model}
To get a better picture of the full kinetics in a streamer discharge and to gain a reference for the rotational and vibrational temperatures from the fitted spectra, a global model for pure nitrogen is set up in PLASIMO~\cite{Dijk2009}.
This model uses the reduced electric field acquired from the 33\,mbar pure nitrogen results, see figure~\ref{33mbarN2Axial} and the Plasimo $E/N$ curve therein, as an input and the data is extrapolated according to the velocity diameter relation from Babaeva and Naidis~\cite{Babaeva2021} for the 66\,mbar streamer shown in figure~\ref{phase_resolved}.
The 33\,mbar pure nitrogen streamer was larger than the monochromator slit entrance, preventing the proper Abel inversion needed for this analysis.
After the maximum, the field decreases to 1/20th of its value for the conductive channel.

Something to note regarding the initial boundary conditions. 
The simulation is setup with initial densities according to their room temperature thermal distributions, except for N$_2^+$ and $n_e$ which are both set to 1$\cdot$10$^8$~m$^{-3}$.
This means the model simulates using the initial conditions of a virgin background gas, in contrast to the experimental discharges which are performed with repetition rates and flushing of the vessel.
 The cross section data from the Lisbon~\cite{Alves2014} database, with some additions from \cite{bilek2019electric}, namely the excitation of the vibrational states of N$_2$(${\rm C}^3 \Pi_{\rm u}$), have been used in the calculations.
 The remainder of the model is largely based on the vibrationally resolved model of \v{S}imek \cite{Simek2018}.
 In order to capture the losses of N$_2$(${\rm C}^3 \Pi_{\rm u}$) processes P632-P635, which represent quenching processes with ${\rm N_2}({\rm X}^1\Sigma_{\rm g}^+)$, are taken from B\'{i}lek \cite{bilek2019electric}.
 Additionally, processes P331 and P364-P367 include ion conversion processes of N$_2^+{({\rm B}^2\Sigma_{\rm u}^+)}$ and N$_2^+{({\rm X}^2\Sigma_{\rm g}^+)}$.
 
\begin{figure}%
    \centering
    \subfloat[]{
        \label{Plasimo_a}{\includegraphics[width=0.40\textwidth]{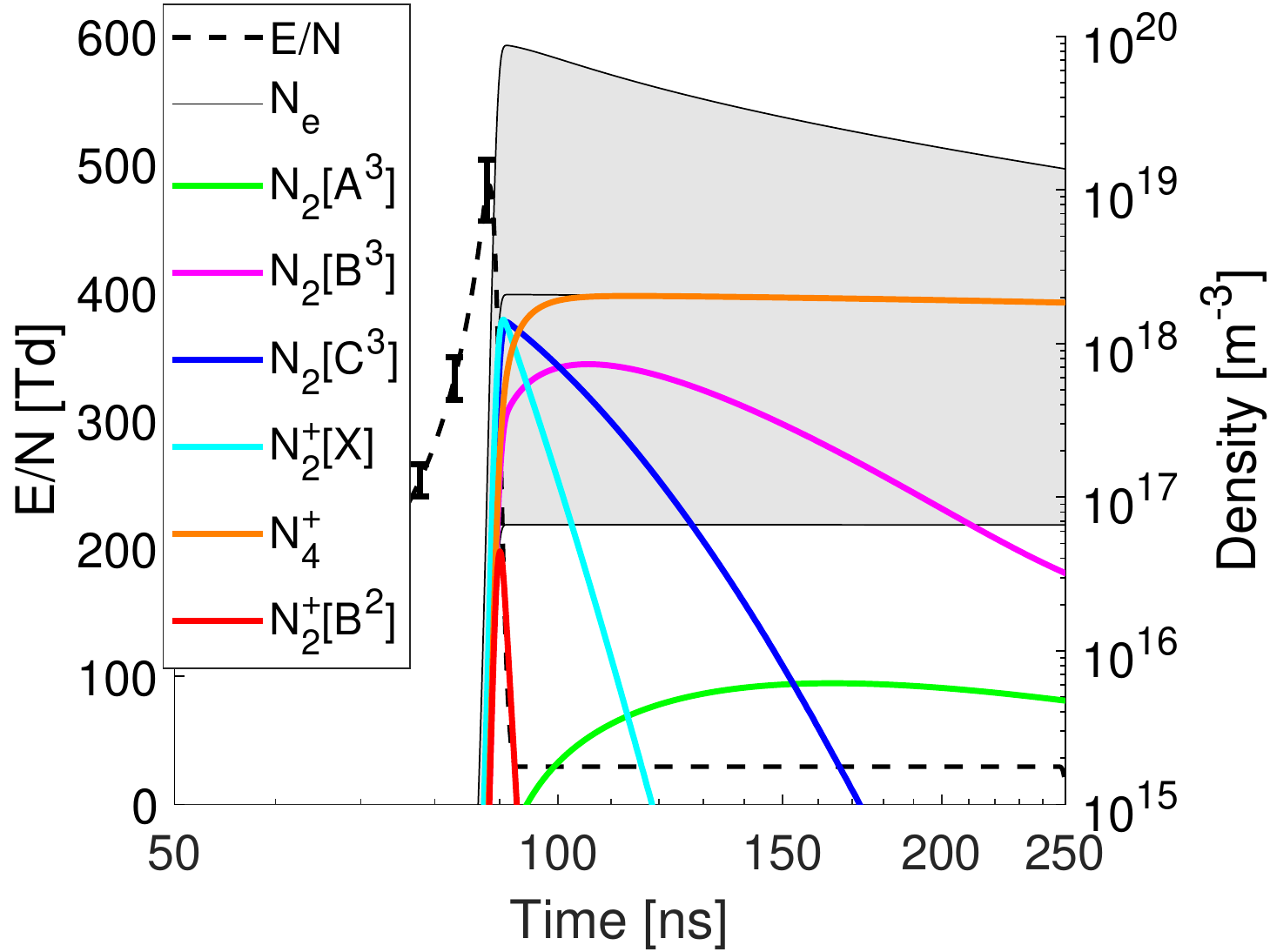} }
    }%
    \quad
    \subfloat[]{
        \label{Plasimo_b}
        \includegraphics[width=0.54\textwidth]{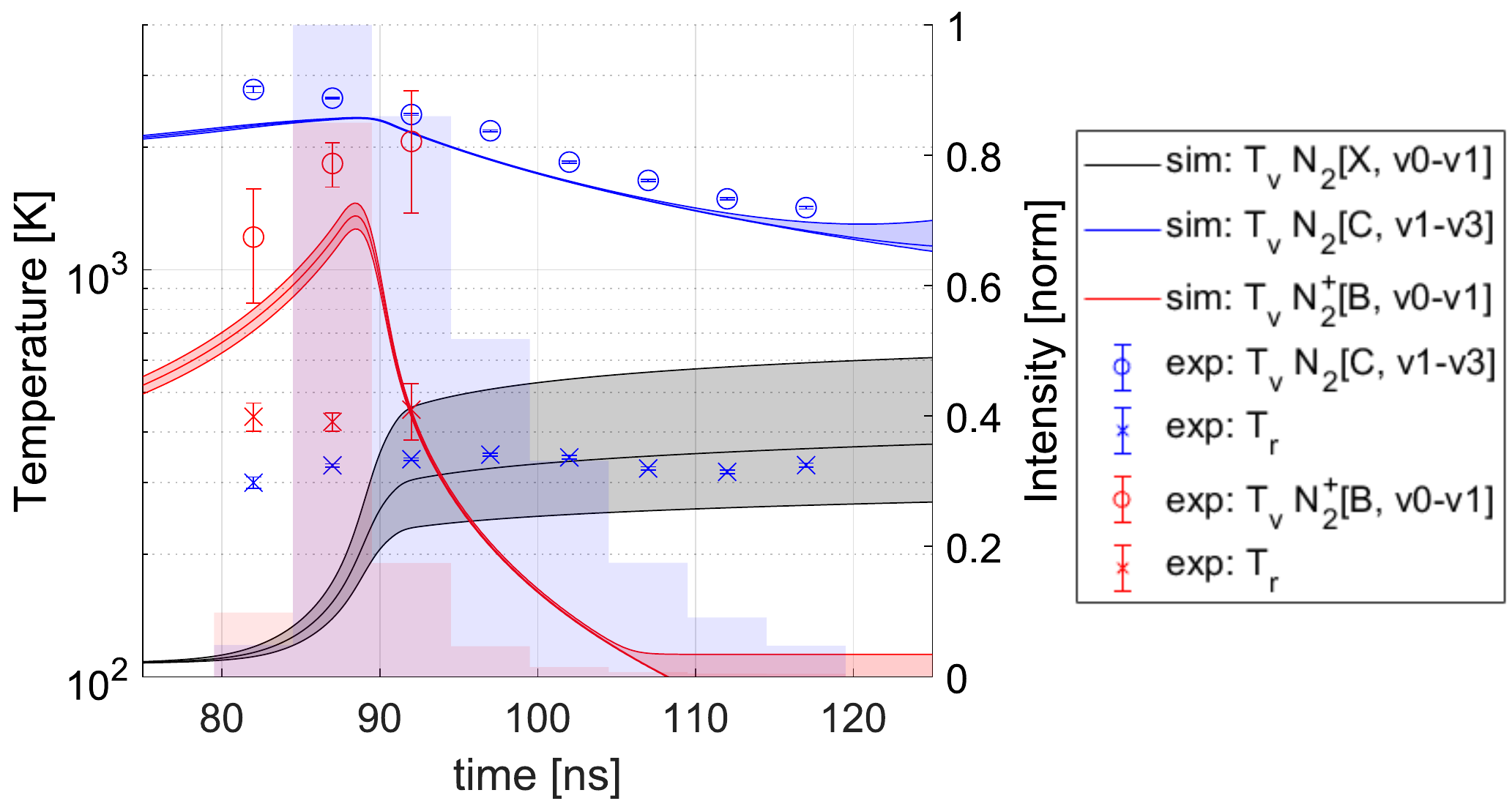}
    }%
    \caption{Global model results for pure nitrogen at 33\,mbar. a) Development of  $E/N$ and the density of some species of interest. The error bars for $E/N$ and shaded region for the electron density denote the $\pm$5\% runs of input $E/N$. b) Fitted SPS and FNS temperatures of both OES experiments and global model, plotted for the pure nitrogen 66\,mbar $E/N$ pulse over the axial center. The outer lines of the shaded regions show the $\pm$5\% runs of input $E/N$. The transparent bar plots show the total FNS and SPS emission and their temporal binning.}%
    \label{Plasimo}%
\end{figure}

In figure~\ref{Plasimo_a} some species densities of interest, i.e., the densities of the first couple of electronic states and the upper states of the FNS and SPS, ${\rm B}^2\Sigma_{\rm u}^+$ and ${\rm C}^3\Pi_{\rm u}$ are plotted as a function of time for times around the space charge layer passage.
The input $E/N$ curve is varied by $\pm$5\% to show the sensitivity of the model, the error bars and shaded regions in figure~\ref{Plasimo} reflect these variations.
A 5\% change in input $E/N$ affects the resulting electron density in the channel by an order of magnitude.
This is because the direct $E/N$ simulation has no self-consistency where the $E/N$ reduces when the gas becomes conductive, like in the streamer simulation of Li et al~\cite{Li2021} .
Figure~\ref{Plasimo_b} shows the temperature fits from these upper level species, for both the simulation and the experiment. 

Experimental values are fit values from the radial center of the Abel inverted 66\,mbar pure nitrogen spectra, shown in figure~\ref{RadialHeatmaps}.
It should be noted that the distribution of vibrational states was investigated by a state-by-state approach.
The Boltzmann plot of vibrational SPS states, where only three out of five vibrational states are tracked (namely the (1-4), (2-5) and (3-6) transitions), was fit to obtain the vibrational temperature of the ${\rm N_2}({\rm C}^3 \Pi_{\rm u})$ state. 
Despite the ${\rm N_2}({\rm C}^3 \Pi_{\rm u})$ vibrational temperature does not have any direct relation to other parameters of the discharge, we used it to perform the comparison of the model results with the experimental data. 
In the high field region, the vibrational distribution is dominantly populated by electron impact~\cite{Dilecce2022}, resulting in a distribution directly related to the Franck-Condon factors. 
The Franck-Condon-like distribution from the zeroth vibrational level of molecular nitrogen ground state is related to the following distribution: $\frac{[{\rm N_2}({\rm C}^3 \Pi_{\rm u}, v=1)]}{[{\rm N_2}({\rm C}^3 \Pi_{\rm u}, v=0)]}:\frac{[{\rm N_2}({\rm C}^3 \Pi_{\rm u}, v=2)]}{[{\rm N_2}({\rm C}^3 \Pi_{\rm u}, v=0)]}:\frac{[{\rm N_2}({\rm C}^3 \Pi_{\rm u}, v=3)]}{[{\rm N_2}({\rm C}^3 \Pi_{\rm u}, v=0)]}$ = 0.56 : 0.19: 0.05 \cite{gilmore1992franck}. 
This distribution corresponds with the vibrational temperature of approximately 2300 Kelvin, and obtaining a similar vibrational temperature in the experiment can signify the dominant electron-impact excitation.
The vibrational distribution corresponding to the electron-impact excitation of the upper state of the FNS is $\frac{[{\rm N_2^+}({\rm B}^2\Sigma_{\rm u}^+, v=0)]}{[{\rm N_2^+}({\rm B}^2\Sigma_{\rm u}^+, v=0)]}:\frac{[{\rm N_2^+}({\rm B}^2\Sigma_{\rm u}^+, v=1)]}{[{\rm N_2^+}({\rm B}^2\Sigma_{\rm u}^+, v=0)]}$ = 1.0 : 0.13 \cite{gilmore1992franck} and corresponds with the vibrational temperature of approximately 1700 Kelvin. 
The experimental vibrational distribution was investigated from the FNS (0-0) and (1-1) transitions.
To contrast these vibrational temperatures of the excited states, the temperature related to the first two vibrational levels of the ground state ${\rm N_2}({\rm X}^1\Sigma_{\rm g}^+)$ is plotted in figure~\ref{Plasimo_b},

Figure \ref{Plasimo_a} shows that ${\rm N_2^+}({\rm B}^2\Sigma_{\rm u}^+)$ is quenched much faster than the neutral excited states (${\rm N_2}({\rm C}^3 \Pi_{\rm u})$, ${\rm N_2}({\rm B}^3 \Pi_{\rm g})$).
Only for a brief moment in time is the electron impact excitation from ${\rm N_2}({\rm X}^1\Sigma_{\rm g}^+)$ strong enough to balance the radiation and quenching losses of the FNS upper state.
The sharp decrease in the ${\rm N_2^+}({\rm B}^2\Sigma_{\rm u}^+)$ vibrational temperature of the model coincides with a decrease in signal strength in the experiment, where the emission intensity is too low to fit the parameters with any certainty, see the overlap between SPS and FNS emissions and the temperatures in figure~\ref{Plasimo_b}.

For the ${\rm N_2}({\rm C}^3 \Pi_{\rm u})$ states, the trends are similar. 
At the beginning (80-90\,ns) the vibrational temperature of order $2.5 \cdot 10^3$\,K is obtained. 
The subsequent decay of the vibrational temperature (t $>$ 90\,ns) is accurately reproduced due to the vibrational relaxation within the ${\rm N_2}({\rm C}^3 \Pi_{\rm u})$ state:

\begin{equation}
{\rm N_2}({\mathrm C}^3 \Pi_{\rm u})_{\nu^\prime=0-4} + {\rm N_2} \longrightarrow {\rm N_2}({\mathrm X})_{\nu^{\prime\prime}>0} + {\rm N_2}({\mathrm C}^3 \Pi_{\rm u})_{w},
\label{vib_rel}
\end{equation}

where $w < v$, meaning the relaxation from the upper vibrational levels to the lower ones.

Regarding the rotational distributions, which are only available from the experimental data, since no rotational resolution/manifold, or gas heating, was included in the model: 
The SPS displays a very thermal distribution but it should be pointed out that such a thermal distribution in the SPS is commonly disturbed by an overpopulation of the higher rotational states which are located in spectral areas where noise is comparable or even stronger than the signal in our measurements (for SPS\,(1-4)\,$\approx$\,394.7-397\,nm), see for example~\cite{Wang_2007}.
Therefore, we expect complete thermalization of rotational states of SPS or at least partial thermalization. 
In both of these cases, the distribution of initial rotational states (J$<\,\sim$\,30) can be used as an estimation of gas temperature with an error up to 9\% according to Bruggeman et al.~\cite{Bruggeman_2014}.
The rotational distribution of the FNS strongly suggests overpopulation of higher states, which points to additional heating of the ions in the strong electric field. 
However, no definite conclusion can be made as the signal to noise ratio in the tail of FNS\,(0-0) is good enough to estimate a rotational temperature but not sufficient for state-by-state fitting.

\setcounter{subfigure}{0}

\section{Conclusions}
Streamers cause transient and spatially localized non-equilibrium heating of the gas, in this work the degree of all is explored.
The excitation and ionization due to electron impact collisions creates highly energetic species in the avalanches, but has a very limited effect on the gas temperature.
Two of these species, N$_2^+$(${\rm B}^2\Sigma_{\rm u}^+$) and N$_2$(${\rm C}^3\Pi_{\rm u}$), when investigated with OES, can characterize the electric field responsible for creating these species in the head of the streamer.

The measured reduced electric fields peak around 500\,Td at the streamer head in both synthetic air and pure nitrogen with the latter being slightly lower.
 In order to obtain the $E/N$ from the measured FNS and SPS intensities we use two approaches; an experimental (according to the modified Paris curve) and a theoretical (as obtained according to Bílek) approach.
Both give very similar results and show the high sensitivity of the method for the $E/N$ in streamer discharges.
The gaseous medium only experiences this field locally for a few nanoseconds and it depends strongly on the diameter and velocity of the streamer, which in turn both depend strongly on pressure, discharge geometry and applied voltage.
In this experiment, the streamers effectively turned the global applied voltage pulse of 8\,kV and 100\,ns into a few nanoseconds long local pulse with an electric field of 4\,kV/cm.


Apart from the determination of $E/N$ by two distinctive approaches and their comparison to models, the spatial and temporal evolution of rotational and vibrational temperatures for N$_2^+$(${\rm B}^2\Sigma_{\rm u}^+$) and N$_2$(${\rm C}^3\Pi_{\rm u})$ is determined for a series of different sub-atmospheric pressures.
The presented global model expands this to the kinetics in the streamer discharge and gives a reference to interpret the experimental results while showing good agreement.


\section{Acknowledgments}
This publication (SD \& SN) is part of the project ‘Let CO$_2$ Spark’ (with project number 15052) of the research programme Open Technology Programme which is (partly) financed by the Dutch Research Council (NWO).
LK and TH were supported both by the Czech Science Foundation under contract no. 21-16391S.
LK is also a holder of Brno Ph.D. Talent Scholarship funded by the Brno City Municipality.
PB was supported by the Czech Science Foundation (Project No. 15-04023S) and by the Strategy AV21 project.

The authors want to thank; Jan Vorac for the MassiveOES software support, Mahdi Azizi for the in-house pulser, Xiaoran Li for the air comparison model results and Eddie van Veldhuizen for proofreading the draft.


\section{Author Contributions Statement}
SD: Experimental setup, experiment execution, data Analysis, main paper text.\\
LK: Data analysis, experiment execution, MassiveOES spectral fitting, main paper text.\\
JJ: Plasimo global model setup and simulations.\\
PB: Making the calibration curve applicable to the experiments and writing the Methods section.\\
SN: Experimental and processing supervision, full text redirection.\\
TH: Processing supervision, full text redirection.


\section{Conflict of Interest Statement}
No conflict of interest was identified.

\section{Contribution to the Field Statement}
Streamers contain a transient non-equilibrium plasma. 
In this work, an experimental setup is designed to create stable and repetitive streamers for high-resolution spectroscopy measurements.
The active species generated in the streamer head are measured and used to calculate the electric field distribution with two different methods. 
These results are compared to literature and simulations, with the latter being specifically adapted to the experiment. 
Additionally, the internal state distribution of the active species in the tail of the streamer is measured for a range of pressures in pure nitrogen and air. 
A global model is then used to interpret the experimental measurements, providing insight into the kinetics and spatiotemporal distribution of energy deposited into the gas.

\clearpage
\bibliographystyle{unsrtnat} 
\section{Tables}
\setlength{\tabcolsep}{24pt}

\begin{table*}[tbh]
\centering
        \caption{\label{IMprocesses}The processes used for the theoretical calculation of the FNS\,(0-0)/SPS\,(2-5) intensity ratio.
	}
\begin{tabular}{ll}

\toprule
No.        &   reaction \cr
\midrule
R1       & ${\rm e} + {\rm N_2}({\rm X}^1 \Sigma^+_{\rm g})_{v=0} \longrightarrow {\rm N^{+}_2}({\rm B}^2 \Sigma_{\rm u}^+)_{v'=0} + 2{\rm e}$ \\
R2        & ${\rm e} + {\rm N_2}({\rm X}^1 \Sigma^+_{\rm g})_{v=0} \longrightarrow {\rm N_2}({\rm C}^3 \Pi_{\rm u})_{v'=2} + {\rm e}$ \\
R3       & $ {\rm N^{+}_2}({\rm B}^2 \Sigma_{\rm u}^+)_{v=0} \longrightarrow {\rm N^{+}_2}(\rm X^2 \Sigma_{\rm g}^+) + h\nu$ (FNS) \\
R4        & $ {\rm N_2}({\rm C}^3 \Pi_{\rm u})_{v=2} \longrightarrow {\rm N_2}({\rm B}^3 \Pi_{\rm g}) + h\nu$ (SPS) \\
R5        & $ {\rm N^{+}_2}({\rm B}^2 \Sigma_{\rm u}^+)_{v=0} + {\rm N_2}({\rm X}^1 \Sigma^+_{\rm g})_{v=0} \longrightarrow {\rm products} $ \\
R6       & ${\rm N^{+}_2}({\rm B}^2 \Sigma_{\rm u}^+)_{v=0} + {\rm O_2}  \longrightarrow {\rm products} $ \\
R7       & ${\rm N_2}({\rm C}^3 \Pi_{\rm u})_{v=2} + {\rm N_2}({\rm X}^1 \Sigma^+_{\rm g})_{v=0} \longrightarrow {\rm products} $ \\
R8       & ${\rm N_2}({\rm C}^3 \Pi_{\rm u})_{v=2} + {\rm O_2} \longrightarrow  {\rm products} $ \\
\toprule
\end{tabular}

\end{table*}

\setlength{\tabcolsep}{6pt}
\begin{table*}[h]
\caption{\label{UQ:intervals} Reaction rate constants and their uncertainties as used in the kinetic model.}
\begin{tabular}{llllll}
\toprule
	no.     &   characteristic constant & values for curve & rate uncertainty & unit & Ref. \\
\midrule
R1  &  $k^{\rm \scriptscriptstyle B}_{\rm \scriptscriptstyle im} $ & $k^{\rm \scriptscriptstyle B}_{\rm \scriptscriptstyle im}(E/N)$(*) & EEDFs x CSs & [cm$^3$s$^{-1}$] &~\cite{vsimek2014optical}\\
R2  &  $k^{\rm \scriptscriptstyle C}_{\rm \scriptscriptstyle im} $ & $k^{\rm \scriptscriptstyle C}_{\rm \scriptscriptstyle im}(E/N)$(*) & EEDFs x CSs & [cm$^3$s$^{-1}$] &~\cite{vsimek2014optical}  \\
R3  &  $\tau_{\rm \scriptscriptstyle 0}^{\rm \scriptscriptstyle B}$ & 65.8 &  (58.0 $\div$ 68.0) & [ns] &~\cite{dilecce2010collision} \\
R4  &  $\tau_{\rm \scriptscriptstyle 0}^{\rm \scriptscriptstyle C}$ & 36.8 &  (35.1 $\div$ 44.1) & [ns] &~\cite{dilecce2006oodr} \\
R5  &  $ k^{\rm \scriptscriptstyle B}_{q, \rm N_2}$ & 8.84 & (6.6 $\div$ 9.8) & $\times 10^{-10}$ [cm$^3$s$^{-1}$] &~\cite{dilecce2010collision} \\
R6  &  $ k^{\rm \scriptscriptstyle B}_{q, \rm O_2}$ & 10.45 & (10.0 $\div$ 11.0) & $\times 10^{-10}$ [cm$^3$s$^{-1}$] &~\cite{dilecce2010collision} \\
R7  &  $ k^{\rm \scriptscriptstyle C}_{q, \rm N_2}$ & 4.4 & (3.7 $\div$ 5.2) & $\times 10^{-11}$ [cm$^3$s$^{-1}$] &~\cite{dilecce2006oodr,kirillov2019intermolecular}\\
R8  &  $ k^{\rm \scriptscriptstyle C}_{q, \rm O_2}$ & 3.7 & (3.2 $\div$ 4.2) & $\times 10^{-10} $[cm$^3$s$^{-1}$] &~\cite{pancheshnyi2000collisional}\\
\toprule
\multicolumn{6}{l}{(*) $k_{\rm im}(x = E/N) = 10^6 \times \exp(A + B \log(x) + C/x + D/x^2 + E/x^3)$, the parameters }\\
\multicolumn{6}{l}{A to E for air and pure nitrogen can be found in~\cite{vsimek2014optical}. Ref. denotes to the source of values for the curve.}  \\
\end{tabular}
\end{table*}

\begin{table}[h]
\centering
\caption{Calibration curves of Paris (multiplied by factor 0.5) and Bílek, both at a pressure of 33\,mbar, described by function $R(E/N) = a \exp { \left( -b \, (E/N)^{-0.5} \right)}$ as plotted in the figure~\ref{cal_curves}. \label{bounds}}\begin{tabular}{lcccc}
\toprule
 &  \multicolumn{2}{l}{Paris curve} & \multicolumn{2}{l}{Bílek curve}\\
\midrule
gas & a & b & a & b\\
\midrule
air &  28.8 & 89.0 & 59.7 & 106.2 \\ 
pure nitrogen & 15.9 & 89.0 & 25.9 & 105.2 \\
\toprule
\end{tabular}

\end{table}

\bibliography{ref}

\end{document}